\documentclass[apj]{emulateapj}
\input psfig.tex
\usepackage{amsmath}
\newcommand{\kms} {$\rm {km}~\rm s^{-1}$}

\newcommand{\mbh} {M$_{\bullet}$}

\begin{document}

\shorttitle{The Dark Matter Halo of M87}
\shortauthors{Murphy, Gebhardt \& Adams}

\title{Galaxy Kinematics with VIRUS-P: The Dark Matter Halo of M87}

\author{Jeremy D. Murphy, Karl Gebhardt and Joshua J. Adams}
\affil{Department of Astronomy, University of Texas at Austin, 1
  University Station C1400, Austin, TX 78712, USA}
\email{murphy@astro.as.utexas.edu}

\begin{abstract}

We present 2-D stellar kinematics of M87 out to R~=~238\arcsec\ taken
with the integral field spectrograph VIRUS-P. We run a large set of
axisymmetric, orbit-based dynamical models and find clear evidence for
a massive dark matter halo. While a logarithmic parameterization for
the dark matter halo is preferred, we do not constrain the dark matter
scale radius for an NFW profile and therefore cannot rule it
out. Our best-fit logarithmic models return an enclosed dark matter
fraction of $17.2^{+5.0}_{-5.0}\%$ within one effective radius (R$_e
\cong 100\arcsec$), rising to $49.4^{+7.2}_{-8.8}\%$ within
2~R$_e$. Existing SAURON data (R~$\le 13\arcsec$), and globular
cluster kinematic data covering  $145\arcsec \le \text{R} \le
554\arcsec$ complete the kinematic coverage to R~=~47~kpc ($\sim
5~\text{R}_e$). At this radial distance the logarithmic dark halo
comprises $85.3^{+2.5}_{-2.4}\%$ of the total enclosed mass of
$5.7^{+1.3}_{-0.9} \times 10^{12}~M_\odot$ making M87 one of
the most massive galaxies in the local universe. Our best-fit
logarithmic dynamical models return a stellar mass-to-light ratio of
$9.1^{+0.2}_{-0.2}$ (V-band), a dark halo circular velocity of
$800^{+75}_{-25}$~\kms, and a dark halo scale radius of
$36^{+7}_{-3}~\text{kpc}$. The stellar M/L, assuming an NFW dark halo,
is well constrained to $8.20^{+0.05}_{-0.10}$ (V-band). The stars in
M87 are found to be radially anisotropic out to R~$\cong
0.5~\text{R}_e$, then isotropic or slightly tangentially anisotropic
to our last stellar data point at R~=~2.4~R$_e$ where the anisotropy
of the stars and globular clusters are in excellent agreement. The
globular clusters then become radially anisotropic in the last two
modeling bins at R~=~3.4~R$_e$ and R~=~4.8~R$_e$. As one of the most
massive galaxies in the local universe, constraints on both the mass
distribution of M87 and anisotropy of its kinematic components
strongly informs our theories of early-type galaxy formation 
and evolution in dense environments.

\end{abstract}


\section{Introduction}\label{intro}

Dark matter is a central component of our current theory of large
scale structure formation. Although the nature of dark matter is
unknown, significant support for this cosmological paradigm comes from
well-motivated physical arguments \citep{gun72, pre74, whi78, fil84} and
the remarkable agreement between N-body simulations of the growth
of structure \citep{fre85, dav85, nfw95,spr05} and observations of the
distribution of galaxies in the local universe \citep{dav82,col01}.

With the increase in computational power seen over the past 30 years,
the spatial resolution of numerical simulations has improved to the
point where individual galaxies are well resolved and their dark
matter halos can be studied in detail \citep{moo98,ghi00,spr08,boy09}.
From the study of both cosmological and galaxy scale simulations,
different parameterizations for a universal dark matter density
profile have emerged. Einasto introduced an early parameterization
\citep{ein65, ein68} based on the S\'{e}rsic profile for the light
distribution in galaxies \citep{ser68}. Other dark matter profile
parameterizations have followed \citep{dub91,nfw97,mor98}. While each
parameterization has found some level of success at describing the
distribution of mass on the scales of galaxy clusters, understanding
the extent and shape of galaxy-sized dark matter halos has met with
mixed success.

Observationally, the study of dark matter halos in spiral galaxies
has outpaced that of ellipticals. This is largely due to the presence
of extended HI discs found in spiral galaxies which provide a clean
dynamical tracer to several effective radii \citep{rub80, van86,
jim03}. Analysis of the circular velocity curves of spiral galaxies
provides some of the strongest evidence for the existence of dark
matter on galaxy scales \citep[see][for a review]{sof01}. Lacking the
extended HI discs seen in spiral galaxies, progress towards
constraining the extent and distribution of dark matter in elliptical
galaxies has proven a greater challenge. Despite this complication,
evidence from gravitational lensing \citep{kee01, man06, san07,
car10}, X-ray gas profiles \citep{hum06, chu08, das10}, planetary
nebulae (PNe) and globular cluster (GC) kinematics \citep{cot01,
  dou07, coc09} and integrated light stellar kinematics \citep{ben94,
  ems04, cap06, tho07a, wei09, for09} has shown that elliptical
galaxies are typically dark matter dominated beyond R~$\sim
1.5$~R$_e$. However, not all galaxies studied show definitive evidence
for the existence of dark matter \citep{ger01, rom03, mon10} and the
best choice of dark halo parametrization remains elusive. These open
questions leave key components of our theories of the growth of
structure, galaxy formation and evolution largely in the dark.

Comparison between the results of various mass estimation methods
return agreement for certain systems and disagreement for
others. \citet{coc09} find good agreement between integrated stellar
light absorption line kinematics and PNe data for a sample of 16
early-type galaxies. Yet in other systems the agreement is poor. In an
analysis of NGC1407, the central elliptical galaxy in a nearby evolved
galaxy group, \citet{rom09} find a discrepancy between the mass
profile determined from GC kinematics and the profile determined by
X-ray gas. For the brightest cluster galaxy in Abell 3827
\citet{car10} determine an enclosed mass via strong lensing that is
$10\times$ higher than the mass determined from X-ray
measurements. Mass discrepancies extend to tracers other than X-ray
gas. Stellar kinematics of NGC821 from \citet{for10} and  NGC3379 from
\citet{wei09} disagree with the PNe measurements of \citet{rom03}.

Each of these methods for estimating mass brings its own set of advantages,
assumptions and limitations. Mass estimates based on X-ray gas have
the advantage of very extended coverage, providing spatial overlap
between the other methods. Yet X-ray gas analysis is limited
to massive galaxies and commonly assumes hydrostatic equilibrium of the
gas. Strong lensing mass estimates avoid this potential pitfall as
it makes no assumptions regarding the energy distribution of the
material within the lens. However, lensing is limited in its
flexibility, as the regions of the universe available for exploration
are dictated by the fixed geometry of the lens and source. Velocity
dispersion measurements from integrated stellar light are effectively
available for all local systems, but require a parameterization of the
dark halo and involve assumptions about the degree of triaxiality of
the system. There is also the challenge of getting stellar kinematics
at large radii where the dark halo comes to dominate the mass. PNe
and GCs have an advantage here as they typically extend to large
radii, yet whether these tracers follow the same dynamical history,
and therefore probe the same formation history as the stars, is not
clear for all systems. A natural approach is to combine various data
sets and methods in order to apply the strengths of one method to
overcome the shortcomings of another. \citet{tre04} and \citet{bol08}
take this approach to good success by using both lensing and stellar
kinematics to break the well known mass-anisotropy degeneracy
\citep{dej92, ger93}.

We focus here on the dark matter distribution in the giant elliptical
galaxy M87, the second-rank galaxy in the Virgo Cluster. M87 has been
extensively studied and a number of groups have made estimates of the
extent of M87's mass profile with a variety of methods. Empirical
formulas, based on the virial theorem and measurements of the central
stellar velocity dispersion, returned some of the earliest mass
estimates for M87 \citep{pov61,bra69,nie84}. \citet{sar78} used
stellar velocity dispersion measurements extending to $R \sim 0.7 R_e$
and the photometry of \citet{you78a} to calculate the mass-to-light
ratio (M/L) as a function of radius and estimate enclosed mass. Since
that time, other mass estimates of M87 using X-ray gas \citep{fab83,
tsa96,mat02,das10} and GC kinematics \citep{huc87,mou87,mer93} have
been made. A comparison of these values to the mass estimate made in
this work is given in \S \ref{encmasscomp}.

The outline of the paper is as follows. In \S \ref{data} we give the
details of the data sets used in our dynamical modeling, with
specifics on the VIRUS-P instrument given in \S \ref{instrument}. An
overview of the data reduction steps is given in \S \ref{overview}, with
the complete details provided in the Appendix. \S \ref{extract} explains the
extraction of the line-of-sight velocity dispersion profile and \S
\ref{template} provides details of the selection of template stars and
their application. In \S \ref{models} we explain the orbit-based
dynamical models. In \S \ref{discuss} we give the results of our
dynamical modeling, with a discussion of our enclosed mass estimates
and a comparison of the logarithmic and NFW halos found in \S
\ref{encmass} and \S \ref{encmasscomp}. We explore possible 
systematics in \S \ref{systematics}. 

We assume a distance to M87 of 17.9~Mpc, corresponding to a scale of
86.5~pc~arcsec$^{-1}$.

\section{Data}\label{data}

We make use of 3 sets of kinematic data to dynamically model M87. At
large radii ($140\arcsec \le \text{R} \le 540\arcsec$) we use
globular cluster kinematics \citep{cot01}. Stellar kinematics from the
SAURON data set \citep{ems04} are used within the central
13\arcsec. New stellar kinematics, taken with VIRUS-P \citep{hil08a},
cover $4\arcsec \le \text{R} \le 238\arcsec$ and add substantially to
the two-dimensional spatial coverage of the galaxy. We provide details
of the stellar surface brightness and globular cluster data in \S
\ref{photo}. The SAURON stellar kinematics are discussed in \S
\ref{sauron}. In \S \ref{vpdata} we describe the observations made
with VIRUS-P. \S \ref{instrument} gives details of the VIRUS-P
spectrograph and \S \ref{datacol} explains the data collection.

\subsection{Photometry and Globular Cluster Kinematics}\label{photo}

The application of both the stellar surface brightness profile and
globular cluster data follow \citet{geb09} (hereafter GT09). The
V-band photometry comes from \citet{kor09}, which is a combination of
HST data from \citet{lau92} and various ground-based
observations. This photometry extends from 0.02\arcsec\ to
2200\arcsec. As the dynamical modeling requires the stellar surface
density, the surface brightness profile is deprojected following the
method of \citet{mag99}. Our globular cluster surface density profile
comes from \citet{mcl99} and is deprojected via a nonparametric
spherical inversion as described in \citet{geb96}. The globular
cluster velocities are reported in \citet{coh97}, \citet{coh00} and
\citet{han01} and compiled in \citet{cot01}. We employ the same cuts
to remove foreground and background contamination as described in
\citet{cot01}. These cuts leave us with 278 globular cluster
velocities which we divide into 11 modeling bins. A line-of-sight
velocity dispersion profile (LOSVD) is then determined from all
globular clusters within one modeling bin as described in GT09.

\subsection{SAURON Stellar Kinematics}\label{sauron}

The SAURON data set provides two-dimensional spatial coverage of M87
out to nearly 40\arcsec\ with superior spatial resolution to
VIRUS-P. We therefore use SAURON kinematics in the central region of
M87. Once the size of the modeling bins makes the SAURON spatial
resolution irrelevant (R~$\ge 8\arcsec$) the VIRUS-P data is used. We
elect to use both SAURON and VIRUS-P kinematics between $8\arcsec \le
\text{R} \le 13\arcsec$ as described in \S \ref{starmodel}.

The publicly available SAURON kinematics are parametrized by the
first 4 coefficients of a Gauss-Hermite polynomial expansion. As our
dynamical modeling fits the full LOSVD rather than its moments we
reconstruct the LOSVD via Monte Carlo simulations based on the errors
provided by SAURON. The details of this reconstruction can be found in
GT09.

\subsection{VIRUS-P Stellar Kinematics}\label{vpdata}

The VIRUS-P data were taken during three separate observing runs over
10 partial nights in January 2008, February 2008 and February
2009. VIRUS-P has no dedicated sky fibers. Therefore, sky nods are
necessary and constitute approximately one-third of our observing
time. All our VIRUS-P data for M87 were acquired through a cadence of
20 minute science exposures bracketed by 5 minute sky nods. We note
that while not having dedicated sky fibers presents issues with
determining the correct level of sky subtraction, sampling the sky
with all 246 fibers allows us to better match the PSF variation from
fiber-to-fiber while not adding substantially to our photon noise. A
discussion of both the advantages and drawbacks of sky nods, and the
details of our sky subtraction method are given in \ref{skysub}.

The VIRUS-P data for M87 consists of 5 pointings extending to
238.0\arcsec\ (20.6~kpc). The pointing placements are shown in Figure
\ref{M87pic}. Exposure times and radial distances for each
pointing are given in Table \ref{exptime}. Ten of the 51 science
exposures were taken under marginal conditions and withheld from the
final data set as they degraded our signal-to-noise (S/N). The
exposure times quoted in Table \ref{exptime} include only the data that
went into the final spectra and subsequent modeling. 

\subsection{The VIRUS-P Instrument}\label{instrument}

The Visible Integral-field Replicable Unit Spectrograph- Prototype
(VIRUS-P), currently deployed on the Harlan J. Smith 2.7~m telescope
at McDonald Observatory \citep{hil08a}, is a prototype for the VIRUS
instrument \citep{hil06}. VIRUS is a replicated, fiber fed
spectrograph currently under development for the Hobby Eberly
Telescope Dark Energy eXperiment (HETDEX) \citep{hil08b}. Originally
designed to conduct a Lyman-alpha emitter survey \citep{ada10,
bla10b}, the VIRUS-P spectrograph is proving an excellent stand-alone
instrument for a wide range of scientific problems
\citep{ada09,bla09,yoa09,bla10a,yoa10}. VIRUS-P is a gimbal-mounted
integral field unit spectrograph composed of 246 optical fibers each
with a 4.1\arcsec\ on-sky diameter. The CCD is a $2048 \times 2048$
back-illuminated Fairchild 3041 detector. The wavelength range for
these observations is 3545--5845~\AA. The fibers are laid out in an
hexagonal array, similar to Densepak \citep{bar88}, with a one-third
fill factor and a large ($107\arcsec \times 107\arcsec$) field of
view. The large fibers and field of view make VIRUS-P an extremely
efficient spectrograph for observing extended, low surface brightness
objects such as the faint outer halos of elliptical
galaxies. Gimbal-mounted directly to the barrel of the telescope,
VIRUS-P maintains a constant gravity vector. Extensive analysis of the 
fiber-to-fiber wavelength solution and fiber spatial PSF has been
conducted and shows negligible evolution over a night. To quantify the
evolution, the location of the centers of the fibers from the twilight
flats taken at the start and end of the night are compared and found
to deviate $\le 0.1$ pixels for all nights. The wavelength solution
determined from the arc lamps taken at dusk and dawn are also
compared. Typical residuals of the wavelength solution to known arc
lines show an rms scatter of $\sim 0.05$~\AA\ for frames taken at the
same time of night. This value of rms scatter does not increase when
arcs from both dusk and dawn are combined. The one exception occurs
with large temperature swings ($\ge10^\circ$~C). Thermal contraction
or expansion of the input and output ends of the fiber bundle can lead
to a change in position and stress pattern on individual
fibers. Localized pressure on a fiber can lead to focal ratio
degradation \citep{cra88,sch03} resulting in changes to the fiber
position and spatial PSF over a night and increased RMS scatter in the
wavelength solution residuals. These effects are subtle, yet can
degrade the quality of our flat-fielding. Therefore, if a temperature
change $\ge10^\circ$~C is seen over a night, the data is split into
two groups and reduced using the calibration frames taken at the
closest temperature. We found this approach was necessary for two
nights in our January 2009 observing run. However, even when a steep
temperature gradient is seen, wavelength and flat-field calibration
frames are necessary only at the start and end of a night's observing.

The median spectral resolution for this VIRUS-P data is 4.75~\AA\ FWHM
as determined from Gaussian fits to strong emission lines in the arc
lamp frames. This resolution corresponds to an instrumental dispersion
(sigma) of $\sim 150$~\kms\ at 4060~\AA\ and $\sim 112$~\kms\ at
5400~\AA. VIRUS-P was refocused between our January/February 2008 and
February 2009 observing runs which led to a non-trivial change
($\Delta$FWHM~$\simeq$~0.5\AA) in the instrumental resolution. As we
frequently combine the spectra from different fibers and different
nights, the change in instrumental resolution is taken into account
when extracting the stellar LOSVDs. The details of how differences in
instrumental resolution are handled can be found in \S \ref{template}.

The assumption of a Gaussian spectral PSF for VIRUS-P proves to be a
good one. To quantify this, we fit Gauss-Hermite coefficients to 4
bright lines in our mercury-cadmium arc lamp frames for all 246
fibers. Over the 4 spectral lines and all fibers the median H3
coefficient is $0.003 \pm 0.013$. The median H4 coefficient is
$0.0003 \pm 0.0117$. Any non-Gaussian line behavior is further
mitigated by the high dispersion of M87, which puts us well above the
instrumental resolution.

\subsection{Data Collection}\label{datacol}

Calibration frames, taken at the start and end of each observing
night, consist of a set of twilight frames, mercury and cadmium arc
lamp frames and bias frames. The twilight frames are used for both
flat-fielding and determining the position and shape of each fiber
profile. The arc lamp frames are used for the wavelength solution and
determination of the instrumental resolution. (see \ref{details} for
more details). The remainder of an observing night involves a sequence
of 5 minute sky nods and 20 minute science frames. The sky nods were
taken 30\arcmin\ off the galaxy center in a region of sky with minimal
field stars and continuum sources and where the galaxy has a surface
brightness of $\mu_{b}~\sim~26.5$ \citep{kor09}. While this position
still includes intracluster light known to extend across much of the
core of the Virgo Cluster \citep{mih05}, the contribution to the total
flux is very low.

\section{Data Reduction Overview}\label{overview}

We provide a brief overview of the data reduction process here, up
through extraction of the kinematics. The extensive details can be
found in the Appendix.

The primary data reduction steps are completed with Vaccine, an
in-house data reduction pipeline developed for VIRUS-P data. The
reduction steps are as follows. All of the science, sky and
calibration frames are overscan subtracted. A master bias is created
by combining all the overscan-subtracted bias frames taken during an
observing run. The arcs and twilight flats are then combined using the
biweight estimator  \citep{bee90}. A $4^{\text{th}}$ order polynomial
is fit to the peaks of each of the 246 fibers for each night. We refer
to this as the fiber \emph{trace}. This polynomial fit is then used on
each science  and sky frame to extract the spectra, fiber by fiber,
within a 5 pixel wide aperture centered around the trace of the
fiber. The wavelength solution is determined for each fiber, and for
each night, based on a $4^{\text{th}}$ order polynomial fit to the
centers of known mercury and cadmium arc lamp lines. The twilight
flats are normalized to remove the solar spectra. These normalized
flats are then used to flatten the science and sky data. Once the
frames are flattened, the neighboring sky frames are appropriately
scaled, combined, and subtracted from the science frames. Cosmic rays
are located and masked from each 20 minute science frame. For the
dynamical modeling, the galaxy is divided into a series of
line-of-sight radial and angular spatial bins. Therefore, fibers that
fall within a spatial bin are combined. This step leaves us with
individual spectra for 88 different spatial bins. Of these 88 spectra,
the 8 central spectra are withheld from the dynamical modeling, as the
SAURON data have superior spatial resolution in the central region. The
next step before the data is ready to model is the determination of
the line-of-sight velocity dispersion profile, described below.

\subsection{Extraction of the LOSVD}\label{extract}

Our method for determination of the line-of-sight velocity dispersion
profile (LOSVD) follows \citet{geb00b} and \citet{pin03}. We give an
overview of the method here.

To begin, an initial guess for a non-parametric LOSVD of the stars is
made. This LOSVD is distributed into 29 velocity  bins and then
convolved with a set of 12 template stars taken from the Indo-US
template library. Selection of the template stars is discussed in \S
\ref{template}. The continuum is divided out of both galaxy and
template spectra prior to fitting. The fitting routine works by
allowing both the weights given to each of the 29 velocity bins and
the weights given to each template star to vary. A parameter to allow
for an adjustment to the overall continuum of the template stars is
also allowed to float. Minimization of the residuals of the fit of the
convolved stellar template spectra to the galaxy spectra is used to
determine the best LOSVD for that given spatial bin and spectral region.

One of the great advantages VIRUS-P provides in the extraction of the
LOSVD and subsequent error estimates is its wide wavelength range
($\sim 2200$~\AA). The wide wavelength coverage allows us to
determine the best LOSVD in five different wavelength regions.
Of the 5 spectral regions sampled (Table \ref{regions}), 4 of the
spectral regions are used in the final modeling. The Ca H~+~K
spectral region (3650--4150~\AA) proves difficult to fit and exhibits
a large systematic offset in all  of the first 4 moments of the LOSVD
from the other 4 spectral regions, likely due to issues with the
continuum division. This region is therefore not included in the
determination of the final LOSVD and error estimate.\footnote{Since
  the completion of the dynamical modeling, the continuum
  normalization issue experienced with the Ca H~+~K region has been
  solved. However, this region is not included in the dynamical models
  as the cost of re-running 1000's of models is prohibitive. We note
  Figure \ref{regcomp} where the Ca H~+~K region is included in the
  analysis of the systematic offset seen in the Mg~$b$ spectral
  region.\label{ftnt1}}
The final LOSVD is created by taking the average of the 4
LOSVDs within each of the 29 velocity bins. Figure \ref{losvdboth}
shows two of the final 88 LOSVDs, with errors, for a bin  at
R~=~24\arcsec\ and R~=~174\arcsec. Overplotted in these figures are
the LOSVDs from the 4 spectral regions used to generate the final
LOSVDs.

A smaller systematic offset was observed for the Mg~$b$ spectral
region (see Figures \ref{regcomp} and \ref{offset}). Yet unlike the Ca
H~+~K offset, which stems from the difficulty in determining the
placement of blue continuum, we believe this offset is inherent to
the Mg~$b$ spectral region and therefore elect to include it in our
final LOSVDs and subsequent modeling. This decision was made as a
trade-off between the $\sim 10\%$ offset in velocity dispersion seen
with this spectral region, and the mitigating effects a
$4^{\text{th}}$ spectral region has on the statistics of the final
LOSVD and uncertainty estimates. We note also that by including the
Mg~$b$ spectral region, our claim of a massive dark matter halo is
strengthened as the direction for the Mg~$b$ offset is towards lower
velocity dispersions. We pick up this discussion in \S \ref{alpha}.

\subsection{Uncertainty Estimates}\label{extract2}

Error estimates for the best-fit LOSVD for each spatial bin are
determined in two ways. The first is made via Monte Carlo
simulations while the second is an empirical method that makes use of
the wide wavelength coverage of VIRUS-P. Then, for each velocity bin
in each LOSVD, the largest of the uncertainties is taken as the
uncertainty for that velocity bin. Both methods are described here.

The first error estimate is made by a Monte Carlo bootstrap method for
each of the 4 spectral regions used in the final LOSVD. The best-fit
convolved LOSVD and set of weighted template stars provide the
starting point for 100 Monte Carlo realizations. Each realization
involves a randomly chosen flux value, drawn from a Gaussian
distribution, for each wavelength. The mean of the Gaussian
distribution is the flux from the best-fit convolved template spectra,
and the standard deviation is set as the mean of the pixel noise
values for that spatial bin as determined in the Vaccine reductions. A
new LOSVD is determined for all 100 realizations and provides a
distribution of values for all 29 velocity bins in the best-fit LOSVD.
The error estimate is the standard deviation of the 100 realizations
within each of the 29 velocity bins. This Monte Carlo simulation is
run on all 4 spectral regions and returns 4 error estimates for each
of the 29 velocity bins in each of the 88 spatial bins. 

The second method for estimating the uncertainty is made by
calculating the standard deviation of the 4 LOSVDs within each of the
29 velocity bins. This error estimate, combined with the 4 from the
Monte Carlo simulations, gives us 5 estimates of the uncertainty
within each of the 29 velocity bins of the LOSVD. The largest
uncertainty at each of these steps is taken as the final uncertainty
used in the dynamical modeling. We note that both the Monte Carlo and
empirical method for determining the uncertainty return similar
results, with the empirical method typically being larger.

\subsection{Stellar Template Library}\label{template}

The template stars used in the extraction of the LOSVD come from the
Indo-US spectral library \citep{val04}. The 12 stars in our final
template library (Table \ref{tlist}) were chosen from an initial list
of 40 stars selected to cover  a range in stellar type and
metallicity. These 12 stars were selected from the initial list as they
returned the lowest residuals when fit to the spectra while still
maintaining a good range in stellar type. As the resolution of the
template stars does not match the instrumental resolution of VIRUS-P,
we must convolve the template  stars with the instrumental resolution
of VIRUS-P. The instrumental resolution varies both between fibers
and, as the instrument was refocused in April 2009, between observing 
runs. A further complication is that spectra from several fibers are
often combined to reach the desired S/N. For overlapping pointings
this combination can involve spectra from opposite ends of the CCD
where the instrumental resolution can be different by as much as
0.7~\AA\ FWHM. For a galaxy like M87, with velocity dispersions around
300~\kms, the error introduced by ignoring this difference is small
($\sim 2\%$). A simple solution, particularly given M87's high
velocity dispersion, would be to convolve all the spectra to the
lowest instrumental resolution. However, as we are interested in
developing data reduction methods to accept all of the galaxies in our
sample, we avoid degrading our resolution to the lowest value in the
following manner.

The instrumental resolution is calculated from Gaussian fits to 8
unblended arc lines from the arc lamp calibration frames taken each
night. As the instrumental resolution values are noisy, particularly
at weaker spectral lines, a small, smoothing boxcar (5 fibers wide) is
run along the spatial direction. Measurements of the focal ratio
degradation of the VIRUS-P fibers show minimal fiber-to-fiber
variation ($\le 2\%$) \citep{mur08}. As focal ratio degradation is the
dominant characteristic of an optical fiber impacting instrumental
resolution, differences in the instrumental resolution across the
spatial direction of the chip are due to optical effects \emph{after}
the light exits the fiber. As resolution changes stemming from optical
effects should be continuous, a boxcar smoothing of the instrumental
resolution values is justified. Differences in the calculated
instrumental resolution from night to night over an observing run are
$\sim 1\%$ and so one instrumental resolution map is made for an
entire observing run. The worst instrumental resolution over our data
set is 5.0~\AA\ FWHM at 4060~\AA\ and 4.4~\AA\ FWHM at 5673~\AA. Once
an estimate of the instrumental resolution for every fiber and for
each observing run is made, the instrumental resolution for each fiber
going into a spatial modeling bin receives a normalized weight based
on the number of exposures going into the final spectra. This approach
gives more weight to fibers that provide more weight to the final
spectra while accounting for differences in instrumental resolution
between fibers and observing runs. Due to M87's high velocity
dispersion, this step amounts to a negligible change in the final
LOSVD.

Initially, we explored using template stars taken with VIRUS-P to avoid
the complications of convolving the template spectra with the
instrumental resolution. The results achieved by this method proved
less robust for two primary reasons. First, the S/N of the Indo-US
spectra is very high. While it is certainly possible to reach this
S/N with VIRUS-P, there are observing time costs to consider. As
using template stars taken with the instrument is effective only if we
are able to fully sample the instrumental resolution across the CCD,
many exposures on the same template star are necessary. The second
limitation is the variety of stellar types available during an
observing run. Although some variety in stellar type and metallicity
is achievable, significant observing time would be lost in attempting
to build up a sufficiently diverse stellar library.

\subsection{Moments of the LOSVDs}

In Figure \ref{moments} we plot the first 4 Gauss-Hermite moments of
the LOSVDs from each of our 88 spatial bins. The colored diamonds
indicate the angular position on the galaxy, with black along the
major axis followed by blue, green, orange and red falling along the
minor axis. For visual clarity, error bars are plotted only for data
along the major axis. The error bars along the other axes are of
comparable size. The vertical dashed lines indicate where the SAURON
kinematics are used over the VIRUS-P data in the dynamical
modeling. Overplotted with open diamonds are moments from the best-fit
logarithmic model at each spatial bin, after averaging over the
angular bins. To minimize visual confusion, the model fits have not 
been plotted in the central region.

\subsection{Systematics in Stellar Kinematics}\label{alpha}

We have found a systematic offset between our measurement of velocity
dispersion when compared to the SAURON data set. The offset is localized
around the Mg~$b$ lines. Figure \ref{offset} plots the velocity
dispersion measured for the combined VIRUS-P wavelength regions used
in the dynamical modeling (red circles) and the velocity dispersion
calculated from just the Mg~$b$ region (green diamonds). Also
plotted are the SAURON velocity dispersions for M87 (black squares). The
SAURON spectral range is 4810--5310~\AA\ and shows a similar offset to
the VIRUS-P Mg~$b$ spectral region. To highlight this difference we
have plotted, in Figure \ref{regcomp}, the VIRUS-P spectra for 5
spectral regions, along with the template fits (red) and calculated
velocity dispersion for each. For this particular spatial bin at
R~=~24.1\arcsec\ the velocity dispersion determined from the Mg~$b$
region is lower than the mean of the other 4 regions by $\sim
30$~\kms. This offset is not an outlier as can be seen in Figure
\ref{offset}. To place a number on this offset we note that the
average velocity dispersion of all the VIRUS-P data points between
$7\arcsec \le \text{R} \le 36\arcsec$ is 301.8~\kms\ when all 4
spectral regions used in the dynamical modeling are included as
described in \S \ref{extract}. The average velocity dispersion when
using just the VIRUS-P Mg~$b$ region over the same spatial range drops
to 281.8~\kms. Over the same spatial region (7\arcsec\ to 36\arcsec)
the average SAURON velocity dispersion is 287.0~\kms.

The cause for this offset is unknown and we do not attempt a detailed
analysis of the offset here. Considering the good agreement between
the SAURON and VIRUS-P results for the Mg~$b$ spectral range, and the
different methods used by both data reduction pipelines to extract
stellar kinematics, the offset is likely intrinsic to this spectral
region. The issues surrounding the Mg~$b$ spectral region for
determination of the velocity dispersion of elliptical galaxies, and
the correlations with both galaxy luminosity and velocity dispersion
are well known \citep{ter81, dre87, wor92, kun01}. \citet{bar02}
compare the velocity dispersion values measured from the Mg~$b$ and
Ca triplet spectral regions for a sample of 33 local galaxies. They
find that the Mg~$b$ region is more sensitive to changes in the
fitting procedure than the Ca triplet region, and exhibits an offset
in velocity dispersion for 48\%\ of the galaxies in their sample, yet
with roughly equal numbers of galaxies showing higher velocity
dispersion values from either one or the other spectral region. Barth
et al. also compare the velocity dispersions of their Mg~$b$ region
calculated when both including and excluding the 5150--5210~\AA\
spectral window. For 32 of their 33 galaxies they find lower velocity
dispersion values when this region is suppressed from the fitting,
with a clear trend towards a larger offset with higher galaxy velocity
dispersion. We have explored this trend by suppressing a similar
spectral region (5150--5220~\AA) from our fitting and find similar
results; velocity dispersion values calculated from the VIRUS-P
spectra where the Mg~$b$ lines are withheld from the fitting are
systematically lower than when these lines are included. However, the
magnitude of our offset is small ($\sim 3$~\kms) and is $\sim 10\%$
of the offset seen by Barth et al. This discrepancy in the magnitude
of the absolute offset value is likely due to differences in the two
kinematic extraction routines used. Interestingly, it is in the
\emph{opposite} direction as naively expected from a comparison of the
SAURON and VIRUS-P Mg~$b$ regions as excluding the Mg~$b$ lines leads
to lower velocity dispersions, not higher ones. This suggests that the
driving force behind the overall offset between the Mg~$b$ spectral
region and the other 4 VIRUS-P spectral regions is not driven
primarily by fits to the Mg~$b$ lines, but rather springs from issue
in fitting that spectral region as a whole. A systematic study of
various kinematic fitting methods over different spectral regions
would be highly illuminating.

\section{Dynamical Models}\label{models}

We employ axisymmetric orbit-based dynamical modeling based on the
idea first presented in \citet{sch79}. The specific details of our
axisymmetric modeling can be found in \citet{geb00b,geb03},
\citet{tho04,tho05} and \citet{sio09}. The models have been shown
accurate to $\sim 15\%$ for recovery of the dark matter halo
parameters \citep{tho05} and stellar M/L \citep{sio09}. Several other
groups have developed their own modeling based on Schwarzschild's
orbit-based method. \citet{dre88} and \citet{rix97} developed an
orbit-based dynamical modeling code for spherical
systems. \citet{van98}, \citet{cre99}, \citet{geb00b} and
\citet{ver02} generalized to axisymmetric systems and \citet{van08}
has developed a triaxial code. Now a number of groups have employed
Schwarzschild's orbit-based method for black hole mass determination
\citep{cre99,ver02,cap02,geb09}, stellar orbital structure and dark
matter content \citep{cre00, geb03, cop04, kra05, cap06, tho07a, for10}. 

We give an outline of the modeling procedure here. First, the galaxy's
surface brightness is deprojected into a three-dimensional luminosity
density. An edge-on inclination is assumed and so the deprojection is
unique. Next, a trial gravitational potential is determined based on
the three-dimensional light distribution and an initial guess for the
stellar M/L, central black hole mass, and the dark matter halo
parameters. Our orbit library is the same as used in GT09. The galaxy
models extend to 2000\arcsec\ over 28 radial, 5 angular, and 15
velocity bins. The gravitational potential and force are calculated on
a grid that is 5 times finer than the grid used to compare to the
data. On average, 25000 orbits are run in the trial gravitational
potential. A superposition of these orbits is created that is both
constrained by the light density profile and is a best match to the
kinematic data. The superposition is accomplished by giving each orbit
a weight as determined by maximizing the function
$\hat{S}$~=~S~-~$\alpha \chi^2$. Here, $S$ is an approximation to the
Boltzmann entropy, $\chi^2$ is the sum of squared residuals between
the model and data LOSVDs (Eqs. 5 and 6), and $\alpha$ is a smoothing
parameter. See \citet{sio09} for a detailed description of both the
creation of the orbit library and determination of the orbit weights.
The steps above are then repeated for a different model, each with a
different stellar M/L, dark halo circular velocity and scale radius.

Three types of models are run. First, we ran a set of dynamical models
with no dark matter halo. As the only free parameter is the stellar
M/L, only 100 models are needed to fully explore the parameter space.
For the cored logarithmic halo (Eq. 3) we ran 6500 dynamical
models. Over 8500 models are run assuming an NFW dark halo profile (Eq
4). For each model a distinct set of orbital weights is used and takes
approximately 1.5 hours of cpu to run. We use the \emph{Lonestar}
computer at the Texas Advanced Computing Center (TACC) at The
University of Texas, Austin to complete all our dynamical modeling.

\subsection{Model Assumptions}

We calculate three types of dynamical models, each assuming a different
mass distribution. First, we consider a mass model for M87 with no
dark matter halo. The mass distribution ($\rho$) for these models
takes the form

\begin{equation}
\rho(r) = \Upsilon \nu(r) + M_\bullet \delta(r)
\end{equation}

\noindent 
where $\Upsilon$ is the stellar M/L, $\nu$ is the three-dimensional
light density and \mbh\ is the black hole mass. As the black hole is
better constrained from GT09 we set our black hole mass to $6.4 \times
10^9~M_\odot$ for all our dynamical models. Gebhardt et al. (2010,
submitted) has refined the black hole mass estimate of M87 to $6.6
(\pm 0.4) \times 10^9~M_\odot$, yet this small change is within our
uncertainties and does not change our results.

Both the second and third sets of dynamical models include a
parameterization for a dark matter halo. The mass distribution then
becomes a sum over each of the mass terms as follows

\begin{equation}
\rho(r) = \Upsilon \nu(r) + M_\bullet \delta(r) + \rho_{\text{DM}}(r)
\end{equation}

\noindent
where the first two terms are the same as in equation 1, and
$\rho_{\text{DM}}(r)$ is the dark matter density term. Two different 
parameterizations for the dark matter halo are explored. The first is
a logarithmic dark matter halo with a density profile as given by

\begin{equation}
\rho_{\text{DM}}(r) \propto v^2_c \frac{2r^2_c + r^2}{(r^2_c + r^2)^2}
\end{equation}

\noindent
The logarithmic halo features a flat central density core of size
$r_c$ and an asymptotically constant circular velocity, $v_c$. The
second dark matter density parameterization is a Navarro-Frenk-White
(NFW) profile \citep{nav96} as given by

\begin{equation}
\rho_{\text{DM}}(r,r_s) \propto \frac{1}{(r/r_s)(1 + r/r_s)^2}
\end{equation}

\noindent
The NFW halo diverges like $r^{-1}$ towards the center and drops as
$r^{-3}$ with radius. The concentration ($c$), scale radius ($r_s$) and
the virial radius ($r_v$) are related via $c = r_v/r_s$. Both dark
matter parameterizations are included in the modeling as described in
\citet{tho05}. 

\subsection{Modeling the Stars and Globular Clusters}\label{starmodel}

The spatial grids used for the modeling are the same as in GT09. The
spatial binning is split into $\text{N}_r = 28$ radial and
$\text{N}_\theta = 5$ angular bins.  Where the model bins are larger
than the SAURON bins, we re-bin the SAURON data. The re-binning is
accomplished by taking the average of all the LOSVDs falling within one
model bin, weighted by their uncertainties. This complication doesn't
arise with the VIRUS-P stellar data as we simply combine all the
spectra from all fibers that fall within a given model bin \emph{before}
extraction of the LOSVD. For the central model bins ($\text{R}
\lesssim 8$\arcsec) we elect to use just SAURON data for its superior
spatial coverage. Between 8\arcsec~$ \le \text{R} \le$~16\arcsec\ we
use both SAURON and VIRUS-P data. We do not combine these data, but
rather send in two LOSVDs independently into the dynamical modeling
routines. A total of N$^{\text{stars}}_{\cal{L}} = 25 + 80$ LOSVDs
(SAURON + VIRUS-P) are used in the modeling, with each stellar LOSVD,
$\cal{L}^\text{stars}$, sampled by $\text{N}_{\text{vel}} = 15$
velocity bins.

To determine the best-fit model, a $\chi^2$ minimization is run in each
trial potential. The $\chi^2$ is calculated as

\begin{equation}
\chi^2_\text{stars} =
\sum_{\text{i}=1}^{\text{N}^{\text{stars}}_{\cal{L}}}
\sum_{\text{j}=1}^{\text{N}_{\text{vel}}}
\left( \frac{\cal{L}^\text{stars}_{\text{ij}} - \cal{L}^\text{model}
_{\text{ij}} \left(\nu \right)} {\Delta \cal{L}^\text{stars}_{\text{ij}}} \right)^2
\end{equation}

\noindent
Here, $\cal{L}^\text{model}_{\text{ij}} (\nu)$ is the i$^{\text{th}}$
model LOSVD in the j$^{\text{th}}$ velocity bin. The orbit model is
forced to reproduce $\nu$, the stellar density, to machine precision. The
residuals between the model and actual set of 105 LOSVDs are minimized
for a single model potential, yielding a single $\chi^2_{\text{stars}}$
value.

As the globular clusters can have a different orbital structure than
the stars, they are treated as a separate kinematic component. The GCs
are handled in a similar fashion as the stars, with the difference
that we employ a deprojected number density for the GCs rather than
the stellar luminosity density as for stars. Both the stars and GCs
are then treated as massless test particles that orbit in a potential
established by the assumed BH mass, stellar M/L and dark halo
parameters. The weighted orbit superposition in each trial potential
is determined by minimizing a similar equation as for the stars,
namely,

\begin{equation}
\chi^2_\text{GC} =
\sum_{\text{i}=1}^{\text{N}^{\text{GC}}_{\cal{L}}}
\sum_{\text{j}=1}^{\text{N}_{\text{vel}}}
\left( \frac{\cal{L}^\text{GC}_{\text{ij}} - \cal{L}^\text{model}
_{\text{ij}} \left(\nu \right)} {\Delta \cal{L}^\text{GC}_{\text{ij}}}\right)^2
\end{equation}

\noindent
where $\cal{L}^{\text{GC}}$ are the N$^{\text{GC}}_{\cal{L}}$~=~11
globular cluster LOSVDs built up from individual GC velocities as
described in \S \ref{extract}. As with the stellar density, $\nu$, the
GC number density is reproduced to machine precision.

\subsection{$\chi^2$ Analysis}\label{chi_analysis}

A $\chi^2$ analysis is used to determine both the best-fit modeling
parameters and their uncertainties. We can rule out a model with no dark
matter with high confidence. The best-fit no dark matter model
returns a stellar M/L$_\text{v}$~=~11.4. However, the $\chi^2$ minimum for
this model is 4898, which is a $\Delta \chi^2 \ge 3571$ increase over
either of the best-fit models including dark matter. We do not discuss
these models further. The best-fit models for both the logarithmic and
NFW halos returns $\chi^2$ minima of 1299.4 and 1310.1 respectively.
The $\Delta \chi^2$ of 10.7 between the two dark matter
parameterizations is statistically significant when comparing the
different constraints we get on the stellar M/L. However, we do not
get a constraint on either of the NFW dark halo parameters,
concentration and scale radius. This is clearly seen in the lower,
right panel of Figure \ref{chi1} where no clear $\chi^2$ minimum for
scale radius is seen out to 350 kpc. As our kinematic data does not
extend beyond 50 kpc we should not expect to get a constraint much
beyond this radial distance. As we do not constrain the NFW dark halo,
we focus here on the logarithmic halo results for our discussion of
the $\chi^2$ analysis, and refer to the NFW results where appropriate.

To select the best-fit dynamical model we analyze the $\chi^2$
values returned from each model run. The $\chi^2$ values plotted in
Figure \ref{chi1} are the additive combination of the $\chi^2$ values
of both stars and GCs, namely, $\chi^2 = \chi^2_{\text{stars}} +
\chi^2_{\text{GC}}$. Figure \ref{chi1} plots these $\chi^2$ values
against the three model parameters for both the logarithmic and NFW
dark halos. Each point gives the $\chi^2$ value from a single
dynamical model. The logarithmic dark halo parameters are plotted on
the left. The solid red line is a cubic spline fit to the lowest
$\chi^2$ values along the parameter space. The dashed blue line shows
the $\chi^2$ minima coming from just the stars. For plotting purposes,
an additive shift of 41.5 has been given to the dashed blue line. As
the shift is additive, the relative $\chi^2$ values are preserved.

On the right side in Figure \ref{chi1} we plot the $\chi^2$ values for
the NFW models. We do not get a constraint on the NFW dark halo scale
radius and concentration parameter. This is evident in the lower-right
panel of Figure \ref{chi1} where the $\chi^2$ minimum runs
unconstrained to $r_s$ values, well beyond the extent of our kinematic
coverage. As the NFW concentration parameter is related to the scale
radius as $c = r_v/r_s$, we also do not constrain this parameter.

A total of 105 stellar LOSVDs and 11 GC LOSVDs are used in the
dynamical modeling. Of the 105 stellar LOSVDs, 25 are determined from
the 4 SAURON moments which provides $25 \times 4 = 100$ parameters.
The 80 VIRUS-P LOSVDs used in the modeling are fit to 15 velocity
bins, giving $80 \times 15 = 1200$ more parameters. The 11 GC LOSVDs
are constructed from 4 parameters, giving another $11 \times 4 = 44$
parameters which totals to 1344 for each dynamical model. The best-fit
dynamical model for a logarithmic halo had a $\chi^2 = 1299.4$, giving
a reduced $\chi^2$ value of 0.97. The $\chi^2$ minimum for the NFW
halo was 1310.1, which gives a similar reduced $\chi^2$ value.

The constraints on stellar M/L come predominately from the stars, yet
the GCs help to constrain the high M/L end as can be seen in the
top-left panel of Figure \ref{chi1}. This result is not surprising.
The GC kinematics constrain the total enclosed mass in the outer
modeling bins; their kinematics strongly influence the resulting dark
matter halo mass. As we assume a constant M/L for the stars, mass not
accounted for in the dark matter halo must get accounted for in the
stars and drive the M/L to higher values. Therefore, kinematics that
constrain the dark halo will also constrain regions of the modeling
where that mass would otherwise wind up, namely higher values for the
stellar M/L.

In the lower-left panel of Figure \ref{chi1} we see a different
influence of the GC kinematics stemming from their extended spatial
coverage. Constraints on higher $r_s$ values come from the GC
kinematics, which extend to 47 kpc. This result is expected, as the
stellar kinematics do not extend out to the dark halo scale radius and
can therefore not influence the modeling. Clearly the GC kinematics
are important for constraining the dark halo parameters, and the good
agreement between the best-fit stars + GC model and the stars-only
model, where the kinematics overlap, is reassuring since it implies
that both large radii stars and globular clusters are in dynamical
equilibrium. Further evidence for equilibrium between the large radii 
stars and GCs is seen in the excellent agreement in their velocity 
anisotropy (see \S \ref{anisotropy} and Figure \ref{intmom}).

The degree to which the GCs help to constrain the dark matter profile
can be seen in another light in Figure \ref{mass2} where we plot the
enclosed dark matter fraction for the logarithmic dark matter
halo. The solid red line shows the dark matter fraction when including 
both GC and stellar kinematics in the analysis. The dashed blue line
comes from an analysis of the stars only. It is clear that kinematics
at large radii are essential to a robust determination of the dark
matter fraction at all radii beyond the central 0.3~R$_e$.

The $\Delta\chi^2 = 1$ range gives us the 68\% confidence bands for
each of the three parameters. For the logarithmic dark halo, the
best-fit stellar M/L is $9.1^{+0.2}_{-0.2}$ (V band). The best-fit
dark matter halo circular velocity is $800^{+75}_{-25}$~\kms, and dark
matter halo scale radius is radius of $36^{+7}_{-3}$~kpc. The NFW dark
halo, while not constrained, still gives a robust estimate of the stellar
M/L of $8.20^{+0.05}_{-0.10}$. The difference in these stellar M/L
values is driven entirely by the shape of the assumed dark halo, and
that the dynamical models work by constraining total enclosed mass. As
the NFW halo allows for a higher central concentration of mass, and
the stellar M/L is assumed constant as a function of radius, mass can
be taken up by the cuspier NFW profile, thus lowering the M/L of these
models.

\section{Discussion}\label{discuss}

The parameters of the dark halo from this paper are different than the
ones presented in GT09 which is also based on a stellar dynamical
analysis. GT09 also fit a cored logarithmic dark matter halo yet find
a circular velocity that is 10\%\ lower and a scale radius that is
60\%\ lower than these results. The reason for the difference is due
to the datasets; the data presented here have substantially improved
kinematic coverage for the stars. The stellar kinematics of GT09 end
at 33\arcsec\ whereas our coverage extends to nearly 240\arcsec. The
GC data is identical between the two papers. The large gap in
kinematic spatial coverage in GT09 between 33\arcsec\ and 140\arcsec\
leads to generally poor constraints on the dark matter halo
parameters. The new VIRUS-P data closes this gap and is therefore more
robust.

\subsection{Enclosed Mass}\label{encmass}

The best-fit dark matter halo parameters for a cored logarithmic
profile returns $800^{+75}_{-25}$~\kms for circular velocity, and 
$36^{+7}_{-3}$~kpc for the scale radius. In terms of enclosed mass,
M87's dark matter halo is one of the largest ever measured for an
individual galaxy. Figure \ref{mass} plots enclosed mass for our
best-fit logarithmic and NFW models. The black and red lines, with
uncertainty, plot total enclosed mass for the logarithmic and NFW
models respectively. The inclusion of a $6.4 \times 10^9~M_\odot$
black hole keeps the total enclosed mass from reaching zero at
R~=~0. The uncertainties are the min/max values for the $4^2 = 16$
dynamical models that explore the parameter limits of our 68\%\
confidence bands. For the uncertainty in the black hole mass we use
the $\pm 0.5 \times 10^9~M_\odot$ values from GT09. The stellar
component is plotted in green (dot-dash) with uncertainties within the
thickness of the lines. The dark matter profiles are plotted in gray
(long dash). The yellow vertical line shows the extent of our
kinematic data. The comparison between the enclosed mass model from
the best-fit logarithmic and NFW halos shows good agreement to the end
of our stellar kinematics. The NFW enclosed mass profile then begins to
diverge to lower total enclosed mass to the end of our kinematic
coverage. We discuss how our results compare to other mass estimates
for M87 in the following section.

\subsection{Comparison to Other Mass Estimates}\label{encmasscomp}

At larger radii \citet{doh09} have measured kinematics of PNe for
M87. They find a dark halo consistent with the one presented here
inside of 500\arcsec, although since their radial range is 400\arcsec\
to 2500\arcsec\ there is not much spatial overlap with our current
data set. At around 600\arcsec\ they find that the mass density begins
to decrease strongly, leading to a truncation of M87's dark halo. At
R~=~1500\arcsec, their outermost radial bin, the PNe dispersion they
measure is $78 \pm 25$~\kms. For the spatial overlap between our work
and theirs (400\arcsec\ to 540\arcsec), where we are now comparing
globular clusters and PNe, the kinematics disagree. Possible reasons
for the disagreement are that the GC kinematics in this region are
poorly measured or that the GCs are not in dynamical equilibrium
(e.g. from a recent merger event). Both \citet{coh00} and
\citet{cot01} find the GC population around M87 shows both chemical
and kinematic evidence for two distinct populations of GCs. Another
possibility is that the PNe measurements are biased in some
way. \citet{doh09} exclude 3 of 8 PNe for their R~=~800\arcsec\ bin as
intracluster planetary nebula and not tracing the potential of
M87. Including these 3 PNe raises their measured dispersion from
139~\kms\ to 247~\kms. Certainly a comparison to either GC or stellar
kinematics at this radial position would be enlightening.

\citet{wu06} estimate the enclosed mass of M87 at 32~kpc (35.1~kpc at
our assumed distance) to be $2.4 (\pm0.6) \times 10^{12}~M_\odot$
using GC kinematics and assuming spherical symmetry. Our mass estimate
of $3.64^{+0.87}_{-0.65} \times 10^{12}~M_\odot$ (logarithmic halo) at
this radial position falls within the uncertainties, yet with an
offset of $\sim 34\%$. \citet{rom01}, using stellar kinematics from
\citet{van94} and \citet{sem96}, and GC kinematics from several
sources, derive an enclosed mass profile for M87 that shows a similar
offset towards lower total mass over the  range 1~R$_e \le$~R~$\le
5$~R$_e$. Within 1~R$_e$ their models diverge to $\sim 50\%$ lower
total mass. This discrepancy may be due to the stellar kinematics used
over this radial range. The stellar kinematics of Sembach \&\ Tonry
exhibit a systematic offset from other data sets for which Romanowsky
\& Kochanek make a correction. The offset between our enclosed mass
and theirs within 1~R$_e$ may be due to this effect or due to the
different modeling assumptions, as Romanowsky et al. assume spherical
symmetry for their modeling. The discrepancy may also come about due
to the increase in spatial coverage the VIRUS-P data affords over
their long-slit spectroscopy.

Comparison of the X-ray mass determination from \citet{das10} to our
mass profile from stars and GCs shows good agreement over the
range 4~kpc~$\le$~R~$\le$~20~kpc, yet diverges elsewhere (see Figure
\ref{others}). At both larger and smaller radii the mass profile from
X-rays is lower than that determined by the stars and GCs. At
R~=~3~kpc the X-ray estimated mass is down by 50 \%\ and at 
R~=~2~kpc the disagreement is $\sim 70$\%. A similar discrepancy is
seen at larger radii. The enclosed mass from X-rays at R~=~47~kpc, our
furthest data point, is lower by 50\%\ than our best-fit value. This
difference is similar to the one seen in NGC4649 \citep{she10}. One
possible explanation for this discrepancy comes from allowing for a
turbulent component in the X-ray gas. A 50\%\ decrease in enclosed
mass can be explained by a $\sim$30\%\ non-gravitational component in
the gas. This amount of of difference is similar to the theoretical
expectation of \citet{bri01} and has been seen in similar systems
\citep{chu10}. More analysis on a wider set of galaxies is necessary
to fully understand the source of these differences.

In Table \ref{masscompare} and Figure \ref{others} we compare enclosed
mass estimates from the literature to this work. Our logarithmic and
NFW halo mass profiles are plotted as in Figure \ref{mass}. Each
colored symbol in Figure \ref{others} indicates the methods employed
to determine the enclosed mass. In general, we find a more massive
dark halo for both our logarithmic and NFW parameterizations, although
our enclosed mass values at various radial positions are not
consistently the highest reported in the literature and appear
consistent with the scatter of the data seen in Figure \ref{others}.

\subsection{Stellar Anisotropy}\label{anisotropy}

The mechanisms by which mass accumulation occurs in galaxies leave
their mark on the distribution function of the stars \citep{lyn67,
val07}. Therefore, mapping the anisotropy of both the stars and GCs
can address questions surrounding galaxy formation history and
evolution. Our orbit-based dynamical modeling return the stellar
orbital structure, which we summarize in Figure \ref{intmom}. Plotted
is the average velocity anisotropy over the 20 angular modeling bins
of both the stars and GCs. The uncertainties are calculated in the
same way as described in Figure \ref{mass} and the text. Within
R~$\simeq 0.5$~R$_e$ the stars show radial anisotropy, then become
mildly tangentially anisotropic to the last stellar data point. The
excellent agreement between the stars and GCs in this region is
indicative of dynamical equilibrium between these two components.
Although we do not conduct a detailed analysis of the anisotropy
of M87 here, comparison of anisotropy maps to N-body simulations can
be highly informative. An example of such an analysis can be found in
\citet{hof10} where the dynamical modeling of NGC4365 by \citet{van08}
is compared N-body simulations. 

\subsection{A Discussion of Systematic Uncertainties}\label{systematics}

Given the high S/N of our data, we pay particular attention to
quantifying systematic uncertainties, since they might be important
for the reported uncertainties. As we are using $\Delta\chi^2$ to
determine the parameter values and uncertainties, if we do not have
proper uncertainty estimates for the kinematics we will bias our final
modeling results. There are three internal consistency checks that
demonstrate that our uncertainties are properly estimated.

First, we estimate LOSVDs and Gauss-Hermite parameters from 4
different wavelength regions. Comparing the standard deviation across
the four regions to the individual uncertainties from the Monte Carlo
simulations provides a consistency check. We find that, in general,
these two uncertainties estimates are consistent. The large wavelength
range of VIRUS-P provides this very important estimate, which includes
both statistical and systematic effects.

Second, the reduced $\chi^2$ for the best-fit dynamical models is near
unity. The $\chi^2$ is measured from the LOSVDs, and we can see the
agreement in the plot of observed and modeled moments (Figure
\ref{moments}). The deviations between the data and the modeling
moments are consistent with the stated uncertainties. While this
consistency does not directly show that systematic effects are not an
issue, it is an indirect confirmation.

Third, when comparing kinematics from datasets taken at different
times we find consistent results within the stated uncertainties. With
the spatial overlap of our pointings \#3 and \#4 we are able to
compare the resulting kinematics from four of our spatial bins when
taken a year apart. We have compared the first four Gauss-Hermite
moments, calculated from our extracted LOSVD, and find that they are
all consistent within their stated uncertainties. These three internal
checks demonstrate control of the measured uncertainties.

Next, we discuss the two areas where systematic effects may be
important: sky subtraction and template mismatch. In order to
determine how the level of sky subtraction affects our extracted
kinematics and subsequent modeling, we explore both over and under
subtraction of each 20 minute science frame. A range of sky
subtraction levels are created and taken through all subsequent data
reductions. A total of 25 different sky subtractions are made on each
science frame, over a range of $\pm 12.5\%$ when compared to equal
exposure times. The details of these reductions are given in the
\emph{Sky Subtraction} section in the Appendix. We then compare the
calculated velocity and velocity dispersion values, taken from the
best-fit LOSVD. This comparison, over all 88 spectra, shows no
systematic offsets in velocity or velocity dispersion for either over
or under subtraction of the night sky. The associated random errors
for this full range of sky subtractions is within our quoted
uncertainty for both velocity and velocity dispersion.

In order to explore possible systematics due to our use of the Indo-US
spectral library, we select the same set of template stars from the
Miles spectral library \citep{san06b} and extract kinematics for all
of our spectra. The two libraries agree very well, with deviations
between the libraries of $\sim 2.5$\kms, well within our quoted
uncertainties for velocity dispersion. In the case of velocity, there
is a slight offset ($\sim 7$\kms) which is due to the lack of a
velocity zero-point between the two libraries. Both of these checks
indicate that our systematics are under control.

\subsection{Next Steps}

This work points the way to several other areas of inquiry. We have
explored two different parameterizations for a dark matter halo, yet
others exist and there is no reason to dismiss any of them. A natural
next step is to rigorously explore several different dark matter halo
parameterizations with the same data sets and modeling methods to
determine which, if any, is favored. This requires us to push the
collection of stellar kinematics to ever larger radii. The amount of
observation time needed to reach to 2.4~R$_e$ with VIRUS-P was not
substantial, and stellar kinematics to 3 and 4 R$_e$ are achievable.
These data would allow for both better constraints on the various dark
matter halo parameters and a comparison with the other dynamical
tracers (i.e. GCs and PNe). As much of our current understanding of
the dark matter halos around elliptical galaxies depends on GC and PNe
kinematics, a robust comparison between each tracer is needed to
explore systematics.

A second avenue of exploration comes from the information contained in
the stellar chemical abundances available through a Lick index
analysis. \citet{gra08} provide a publicly available tool that is
well-suited for this work. How elliptical galaxies formed and whether
their stars were formed in situ or accreted over time requires both a
dynamical and chemical analysis \citep{gra10}. The chemical composition
of GCs at large radii have been studied \citep{coh00, cot01}, and a
detailed comparison of both the kinematics and chemistry of both GCs
and stars at the same radial position should prove immensely fruitful.

Finally, work towards a more complete and uniform sample of massive
elliptical galaxies, both 1$^{\text{st}}$ and 2$^{\text{nd}}$ rank
galaxies, and equally massive field ellipticals (e.g. NGC1600) is
needed to explore the influence of environment on dark matter
halos. Several groups have made significant progress towards this end,
yet the data sets that involve both 2D spatial coverage at both small
and large need to be expanded.

\section{Acknowledgments}

The authors would like to thank The Cynthia and George Mitchell
Foundation for their generous support which made the fabrication of
VIRUS-P possible. J.D.M. would like to thank the members of his
research committee: Michele Cappellari, Gary J. Hill, John Kormendy,
Phillip J. MacQueen, and Milo$\breve{\text{s}}$
Milosavljevi$\acute{\text{c}}$ for all the help given. J.D.M. also
thanks Guillermo Blanc, Ross E. Falcon, and Remco van den Bosch for
their insight and many fruitful discussions. The authors also thank
Dave Doss, Kevin Meyer, Brian Roman, John Kuehne and all of the staff
at McDonald Observatory who helped immensely with the commissioning of
VIRUS-P and the successful collection of this data. This research is
partially based on data from the publicly available SAURON Archive,
which aided substantially in the modeling. The authors thank Payel Das
and her collaborators for their willingness to share their X-ray data
and analysis on M87. This work would not have been possible without
the resources of the Texas Advanced Computing Center at the University
of Texas, Austin where all of the dynamical modeling was run.

\clearpage
\bibliographystyle{apj}
\bibliography{all}

\clearpage
\appendix

\section{Data Reduction}\label{reduc0}

The reduction of integral field spectroscopy (IFS) data involves
numerous issues not faced in the reduction of traditional long-slit
data \citep{bar88,par90,wys92,bar93,lis94,wat98}. Each fiber exhibits
its own character, with variations in spatial PSF, transmission and
focal ratio degradation \citep{avi88,ram88,ber04,mur08}. Despite these
complications, many groups have developed robust and versatile
pipelines for the reduction of IFS data\citep{bac01, zan05, tur06,
  san06, san10}.

This paper is the first in a series and establishes our principle
methods of data reduction. For this reason we give a detailed
description of each step in the data reduction process. In the
description below, the term ``spectral'' is used in indicate the
wavelength or X-direction on the CCD. The term ``spatial'' is used for
the cross-dispersion or Y-direction.

\subsection{Reduction Details}\label{details}

\begin{figure*}[t]
  \centerline{\psfig{file=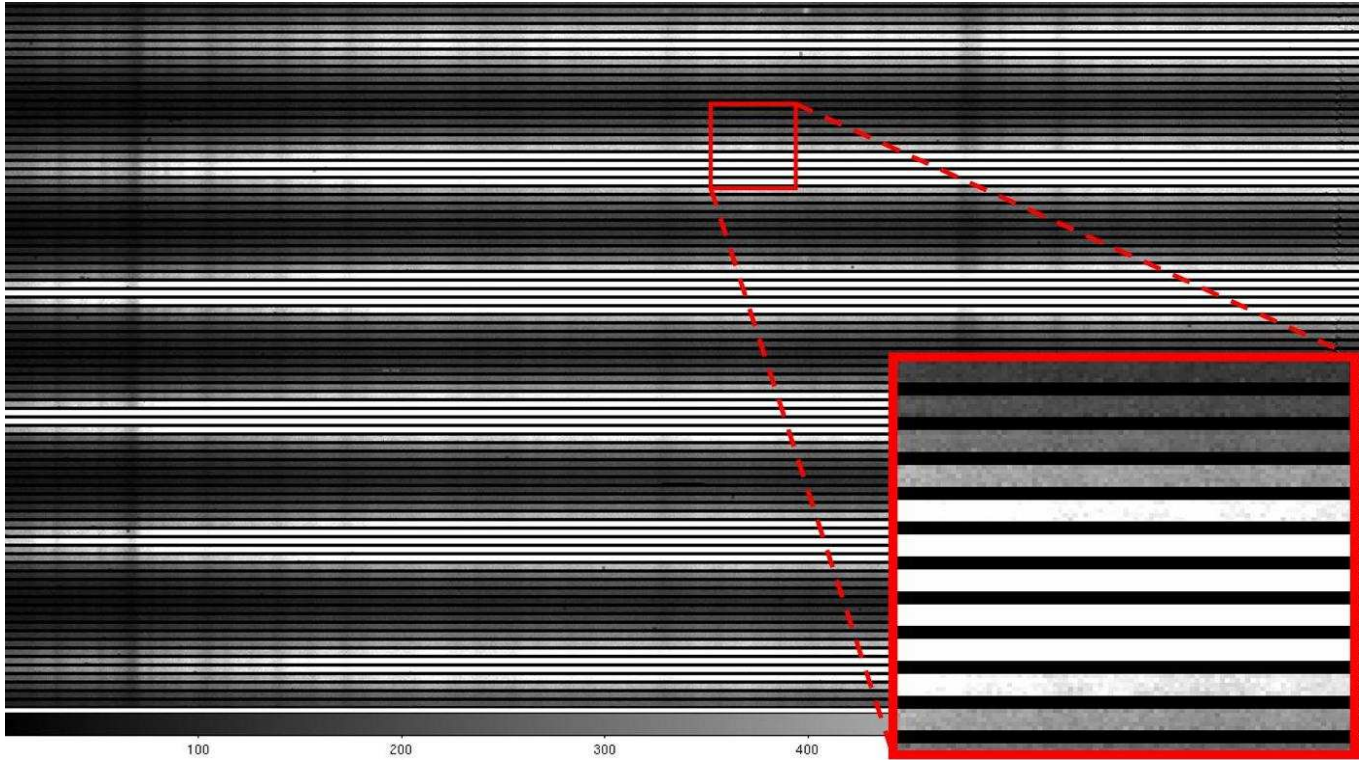,width=18cm}}
  \figcaption[M87cent.ps]{An image of VIRUS-P data from a 20 minute
    central pointing on M87 (\#3 in Figure \ref{M87pic}) after all
    preliminary data reduction is complete. Just the central $\sim 90$
    fibers are shown. The inset shows a close-up of 11 fibers
    extracted over a 5 pixel-wide aperture. Residuals from the
    5577~\AA\ sky line subtraction can be seen to the far right. The
    strong absorption feature seen on the far left is the G-band
    ($\sim 4310$~\AA\ rest frame). The weaker absorption band near the
    center, just to the left of the small box, is H$_\beta$ ($\sim
    4860$~\AA) and the strong, wide feature towards the right is the
    Mg~$b$ region ($\sim 5167$~to~5183~\AA).
    \label{M87data}}
\end{figure*}

\begin{figure*}
  \centerline{\psfig{file=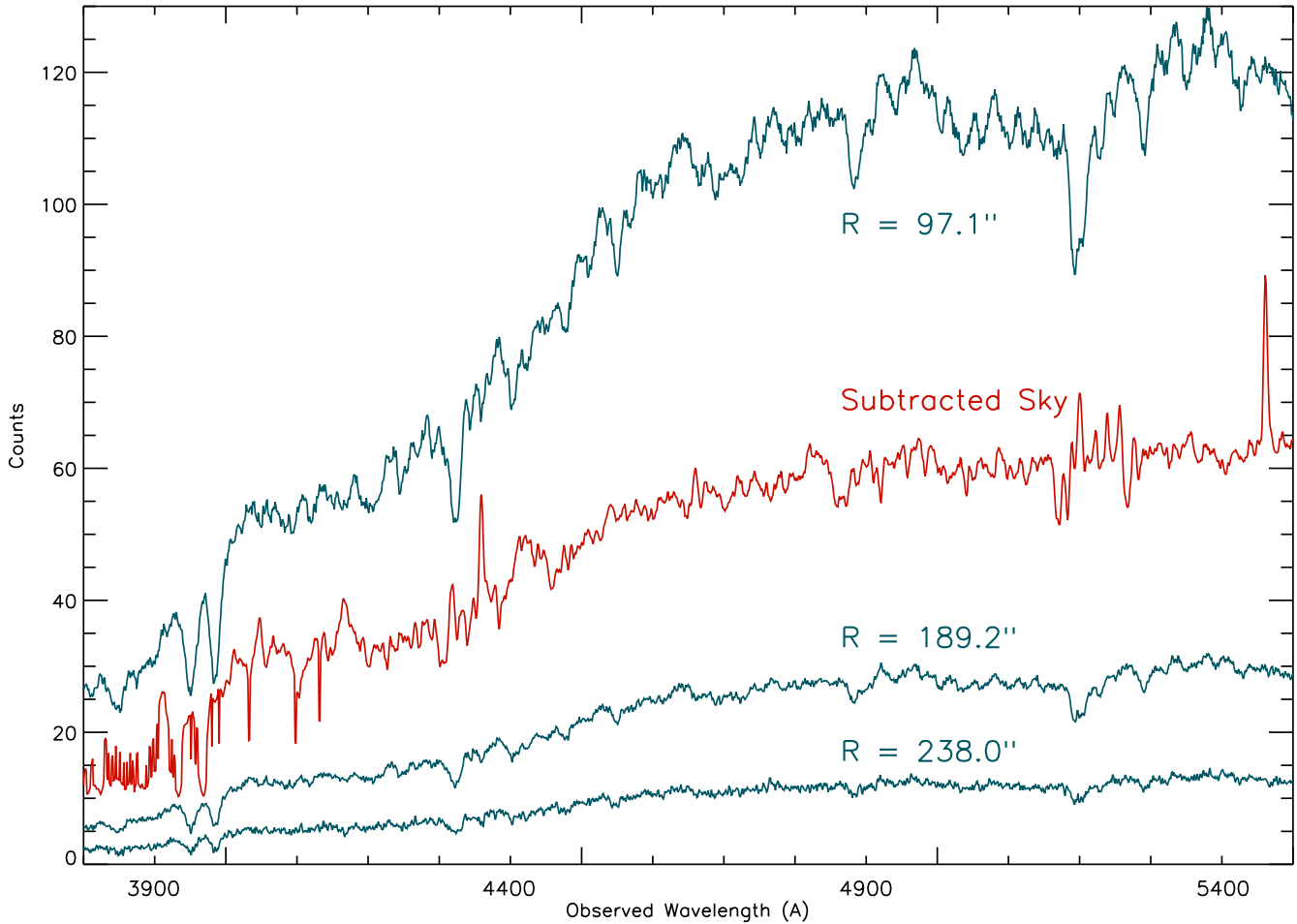,width=20cm,angle=0}}
  \figcaption[data_v_sky.ps]{Spectra from three of the 88 spatial bins
    located at R~=~97.1\arcsec, 189.2\arcsec\ and 238.0\arcsec. The
    counts are a biweight combination of the CCD counts in ADU, after
    sky subtraction, over 20 fibers, 72 fibers and 38 fibers
    respectively. The typical level of the night sky is shown for 
    comparison. The night sky subtracted from the most distant
    pointing at R~=~238\arcsec\ is $\sim 5$ times brighter than the
    galaxy.
    \label{datavsky}}
\end{figure*}

The preliminary data reduction uses Vaccine \citep{ada10}, an in-house
pipeline developed for reduction of VIRUS-P data. We give a full
account of the Vaccine data reduction steps here. First, the overscan
and bias are subtracted from all frames. The CCD is very clean,
therefore no masking of bad pixels is conducted. Next, the twilight
flats and arc lamp frames are combined with the biweight estimator
\citep{bee90}. The biweight is used at several steps in the reduction
process.

Due to issues of instrumental alignment and the inherent limitations
of all optical elements, curvature in both the spatial and spectral
directions on the CCD is unavoidable. The curvature along the spatial
direction is handled by allowing each fiber to have its own
wavelength solution. In order to correct for the curvature along the
spectral direction the twilight flats are used to locate the centers
of each fiber and determine the fiber \emph{trace}. To accomplish
this, a 21 pixel-wide boxcar is run along a single fiber of the
twilight flat in the spectral direction. The fiber profile in the
spatial direction is super-sampled with the boxcar and fit with a
Gaussian profile to determine the center of the fiber at each pixel
step in the spectral direction. This boxcar method effectively smooths
over fiber spatial profile variation due to solar absorption features
and pixel-to-pixel flat-field variation while giving a robust estimate
of the location of the center of the Gaussian profile.  As the
curvature in the spectral direction is not extreme ($\sim \Delta 5$
pixels from the center of the CCD to the edge of the $2048 \times
2048$ chip) a 21 pixel-wide boxcar smoothing is justified. The
location of the centers of all the Gaussian profiles for a single
fiber are then fit with a 4th order polynomial. The polynomial fit
becomes the \emph{trace} of the fiber. The steps described above are
repeated for all 246 fibers.

Once the fiber trace is determined for all 246 fibers, the fiber
profile is extracted, fiber by fiber, over a 5 pixel-wide aperture.
Figure \ref{M87data} shows an image of the central $\sim 90$ fibers of
a typical frame after extraction. As the fiber centroid moves from one
row of pixels to the next the 5 pixel extraction aperture follows. By
allowing the extraction aperture to make discrete steps between rows
of pixels we avoid interpolation of the data. There are two advantages
to not resampling the data at this step. First, we avoid introducing
the correlated noise inherent to interpolation and can therefore carry
accurate pixel-to-pixel noise calculations through the final step of
the Vaccine reductions. This is helpful as a proper S/N calculation
is necessary for the Monte Carlo error estimations made later in the
reductions (\S \ref{extract2}). Second, interpolation can artificially
broaden the spectra, and while the dispersion of M87 is well above the
instrumental dispersion for all our pointings, this should not be
assumed a~priori.

\begin{figure*}
\centerline{\psfig{file=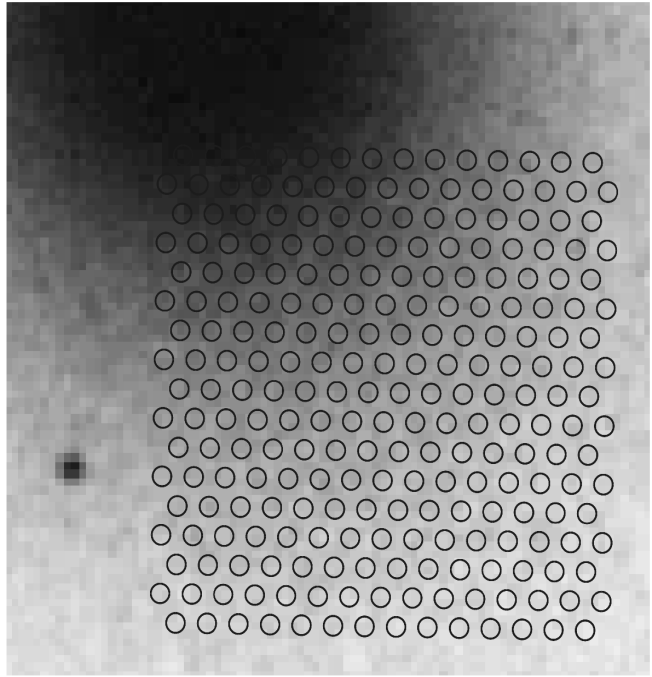,width=8.6cm,height=9.0cm,angle=0}}
\vskip 10pt
\figcaption[fiber_pos.ps]{The relative positions of the 246 fibers
  comprising pointing \#4. The fibers are aligned in a hexagonal array
  with a one-third fill factor. Each fiber has a 4.1\arcsec\ on-sky
  diameter.
\label{fiber_pos}}
\end{figure*}

The typical FWHM of a fiber profile along the spatial direction is
$\sim 4$ pixels with an average spacing of $\sim 8$ pixels between the
centers of adjacent fibers. We have measured the cross-talk between
fibers to be $\le 1\%$ over a 5 pixel-wide aperture. The fiber
position on-sky is mapped onto the CCD from left-to-right and
top-to-bottom (Figure \ref{fiber_pos}). Therefore, neighboring fibers
on the CCD are typically neighboring fibers on the sky and, as
neighboring fibers are often combined to reach the desired S/N, the
effect of cross-talk is further mitigated. We explored extracting over
a 7 pixel-wide aperture and compared the final S/N of both
extractions. Due to the low level of signal in the edges of the 7
pixel aperture, the 5 pixel aperture returns better S/N and is used
for all VIRUS-P data presented here.

Mercury and cadmium arc lamp frames are used for wavelength
calibration and afford 8 unblended and well-spaced emission lines over
our wavelength range. The wavelength of each emission line has been
confirmed using the Robert G. Tull Coud$\acute{\text{e}}$ spectrograph
on the 2.7~m telescope in the R~=~60k set up. The wavelength solution
for each fiber is determined as follows. For an individual fiber, each
emission line is fit with a Gaussian profile to determine its
center.\footnote{We have characterized the spectral PSF of the VIRUS-P
  instrument and find it to be very nearly Gaussian. To quantify this,
  we fit Gauss-Hermite coefficients to 4 bright lines in our
  mercury-cadmium arc lamp frames. Over the 4 spectral lines and all
  246 fibers the median H3 coefficient is $0.003 \pm 0.013$, while the
  median H4 coefficient is $0.0003 \pm 0.0117$.} A $4^{\text{th}}$
order polynomial is fit to the centers of each emission line and the
residuals between the actual wavelength and fit wavelength are
minimized. The cadmium 3611.3~\AA\ line and the mercury 5769.6~\AA\
line are near the blue and red edges of our wavelength range and
minimize the amount of extrapolation needed at the edges of the
polynomial fit. Typical rms residuals of the polynomial fit are $\le
0.07$~\AA\ or $\le 4.4$~\kms\ at 4800~\AA\ (FWHM) . Comparison of the
wavelength solution from the arc lamps taken at the start and end of
the night show differences well below the noise of the fit. We find
that small linear shifts in the spectral direction of the fiber,
possibly due to thermal variations at the output end of the fibers,
can occur on the timescales of an hour. To account for this shift, a
correction is made to the $0^{\text{th}}$ order term of the wavelength
solution based on the change in position of the bright 5577.34~\AA\
sky line. The average of this correction over all frames was 0.13~\AA\
with a standard deviation of 0.11~\AA. A heliocentric correction is
made to each frame. For our February data this correction had a mean
of 19.3~\kms\ and a standard deviation of 0.7~\kms. For our January
data the mean heliocentric correction is 27.6~\kms\ with a standard
deviation of 0.2~\kms.

The next reduction step involves creating a flat-field frame from the
twilight flats. There are four different pieces of information
combined in the twilight flats: pixel-to-pixel variation,
fiber-to-fiber throughput variation, fiber cross-dispersion profile
shape, and the twilight sky spectrum. The first three are aspects of
the flat-field we want to preserve while the twilight sky spectrum must be
removed. Our approach is to construct a model of the twilight
sky, free of flat-field effects, and then divide this model out of the
original twilight frame. To generate a model of the night sky we use
a method similar to \citet{kel03} for IFS sky subtraction. We outline
our method here.

To model the twilight sky a 51 fiber-wide boxcar is run along the
spatial direction. As each fiber has a slightly different wavelength
solution, a B-spline interpolation \citep{die93} of the pixel's
wavelength is made, based on the wavelength solution determined by the
arc lamp polynomial fits. By employing a B-spline interpolation, we
are not limited by the pixelization of the wavelength solution. The 51
fiber-wide boxcar and 5 pixel-wide extraction aperture provides $51
\times 5 = 255$ estimates of the twilight flux at a given
wavelength. Both pixel-to-pixel, fiber throughput, and fiber profile
shape vary on scales much smaller than the size of the boxcar and are
thus smoothed out. What remains is a model of the twilight sky with
flat-fielding effects removed. The model is then divided out of the
original twilight flat, leaving pixel-to-pixel, fiber throughput, and
fiber profile shape intact. We attempted to avoid these complications
through the use of dome flats, yet the intensity of the available dome
lamps below $\sim 4000$~\AA\ is too low to determine an accurate fiber
trace. Even if an acceptable light source was available, there is
another issue with dome flats. It has been shown that the input
acceptance angle is preserved through optical fibers and that focal
ratio degradation is dependent on this angle \citep{car94b, mur08}. As
light entering the fibers from a dome lamp is not collimated, there is
a concern that the fiber cross-dispersion profile is not being
properly quantified with the use of dome flats. Twilight flats avoid
both of these issues.

Initially, the wavelength solution is estimated from un-flattened
arc frames. This can lead to errors in the wavelength solution
when an arc line falls on top of a flat-field feature. Therefore,
the determination of the wavelength solution and subsequent derivation
of the flat-field frame is iterative. The arc lamp frames are
flattened, and the wavelength solution is recalculated. As the
flat-fielding procedure relies on the wavelength solution, a new
flat-field, based on the new wavelength solution, is also made. We
find this iteration leaves the wavelength solution for most fibers
unaffected, yet can improve the residuals by $\sim 0.05$~\AA\ for a
handful of fibers where one or more of the arc lines used for the
wavelength solution fall on strong flat-field features. A single
iteration is all that is required.

The flat-field frame captures the pixel-to-pixel, fiber-to-fiber and
fiber profile shape variation for each fiber at very high S/N.
However, due to thermal effects over an observing night, the science
and sky frames can exhibit a shift in the position of the fiber
profile. This shift manifests as a breathing mode and can reach up to
a 0.3 pixel shift in the center of a fiber when the temperature
gradient over the night is steep. A shift in the center of a fiber
will lead to large flat-fielding errors if not accounted for. To
correct for this effect we have developed a  heuristic solution. The
general idea is to measure the offset over a subset of fibers, then
create a unique flat for each science and sky frame based on the
master flat for that night. We will refer to this frame as the
\emph{science flat} and it is generated as follows. For each fiber in
each science frame the  difference between the fiber center of the
master flat and science frame is calculated at all 2048 pixel
positions. The median of these values for each fiber is taken, then
smoothed with a 12 fiber-wide boxcar. As the breathing mode is smooth
and continuous, and the signal in the science and sky frames can be
quite low, this level of smoothing is both required and justified. 

The fiber profiles have shapes that deviate slightly from any simple
parameterization. For all resamplings we employ a sinc interpolation,
chosen for its non-parametric properties. Simply resampling each
fiber's flat will properly capture the shift in fiber position, but
will improperly capture the pixel-to-pixel features. Therefore we run
a 81 pixel-wide boxcar along the dispersion direction to isolate the
fiber profile from the pixel-to-pixel variation. The original flat
containing the proper pixel-to-pixel map and the resampled fiber
profile are combined to form the final science flat. As sinc
interpolation is not flux-conserving, the science flat is renormalized
to match the total counts in the original science frame. Once this
unique flat is applied, we are left with science and sky frames that
have been extracted, wavelength calibrated and flattened. The next
step is sky subtraction.

\subsection{Sky Subtraction}\label{skysub}

\begin{figure*}
  \centerline{\psfig{file=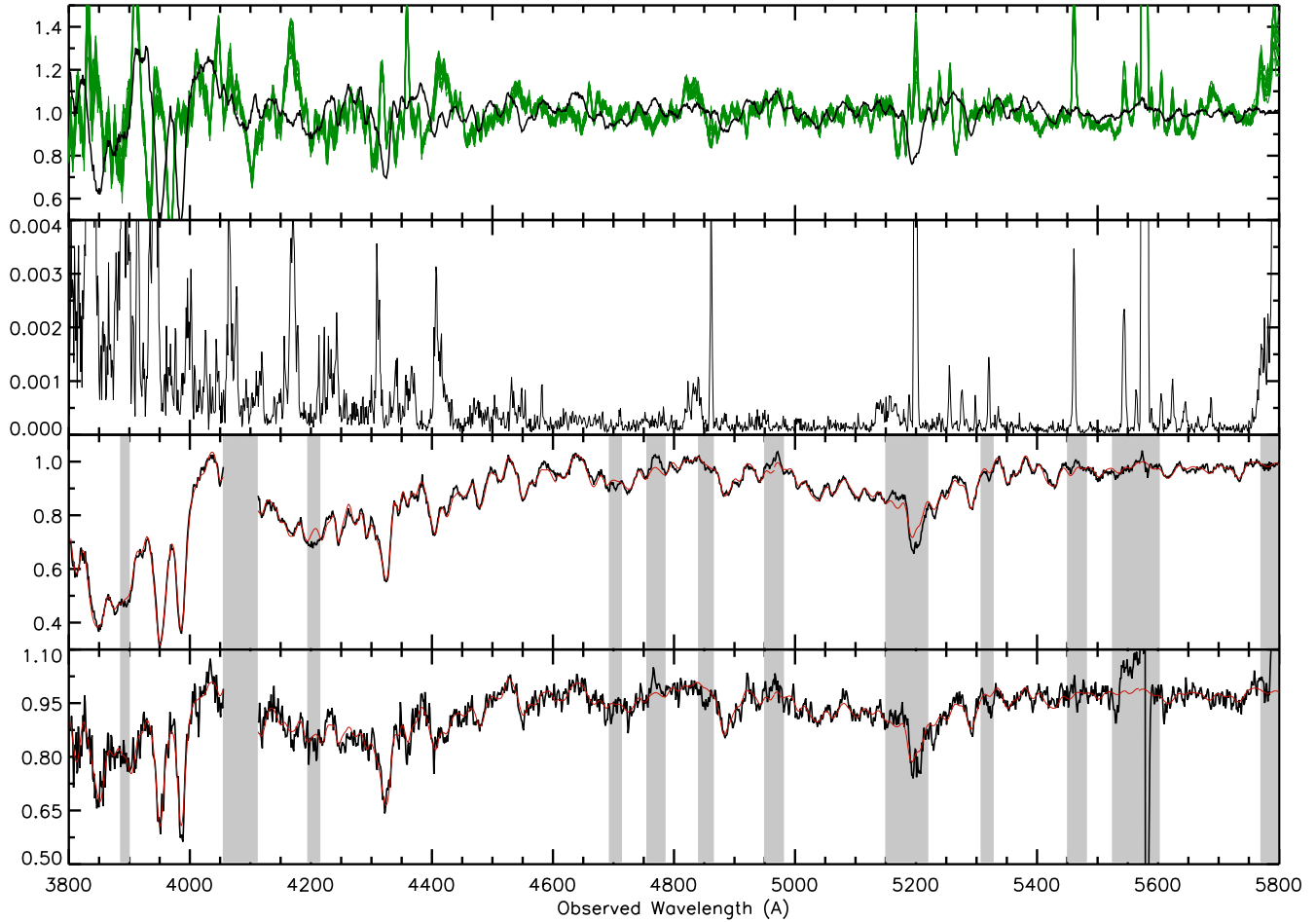,width=19cm,angle=0}}
  \figcaption[sky.ps]{\emph{Top:} The final VIRUS-P spectra for a
  spatial bin at R~=~60\arcsec. The black spectra is the biweight
  combination of 6 fibers over 12 exposures. Overplotted in green are
  the 12 sky spectra subtracted from each exposure prior to the biweight
  combination. The continuum has been normalized here for plotting
  purposes. \emph{Middle Top:} The variance of the 12 sky spectra
  shown in the top panel. Note that while certain high variance
  regions are associated with night sky lines (e.g. 5200~\AA), others
  are not associated with any strong night feature (e.g. 4860~\AA).
  These variance plots are calculated for each spatial bin and used
  to help determine locations where the sky evolves on short time
  scales. \emph{Lower 2 Panels:} Shown in black are the final
  spectra for two spatial bins at R~=~13.2\arcsec\ (upper) and
  R~=~222.0\arcsec\ (lower). Overplotted in red is the best-fit
  stellar template spectra. The gray areas in both lower figures indicate
  spectral regions that are suppressed when completing the kinematic
  extraction due to either issues with sky subtraction or template
  mismatch. We note that while these spectral regions are not fit
  when extracting the final stellar kinematics, the difference in the
  final LOSVD when the smaller regions ($\Delta \lambda \lesssim 50 
  \AA$) are not masked is minimal. Table \ref{regions} shows how 
  the spectra are split into 5 spectral regions prior to stellar
  template fitting.
\label{sky}}
\end{figure*}

Figure \ref{datavsky} plots M87 spectra from 3 different spatial
bins. An estimate of a typical night sky spectra is shown for
comparison. The spectra are shown in CCD counts and the relative flux
between the spectra has been preserved. At 1~R$_e$ the galaxy is still
brighter than the night sky by about a factor of 2. By 2~R$_e$ the
night sky is now more than twice as bright as the galaxy, and in our
furthest spatial bin the sky is $\sim 5$ times brighter. A careful
handling of sky subtraction at these low surface brightnesses is
important, and we discuss our method in detail here. VIRUS-P does not
have sky fibers and so sky nods are necessary. Lacking sky fibers has
obvious drawbacks as we sample the sky at a different point in time,
and loose science observing time to sky nods. Despite these
disadvantages there are benefits to sky nods. One clear advantage to
sky nods comes from the much improved noise statistics we get from
sampling the sky with all 246 fibers. As our sky subtraction is done
with a model of the night sky (described below) the addition of noise
from the night sky estimate is reduced by $\sqrt{\text{N}}$, where N
is the number of fibers. In contrast, Densepak and Sparsepak on the
WIYN telescope dedicate 4.4\%\ and 8.5\%\ of their CCD to sky fibers,
respectively, \citep{bar88, ber04} while the SAURON instrument
dedicates just over 10\%\ of its lenslets to estimates of the sky
\citep{bac01}. Sky nods also avoid the risk of cross-talk between the
science and sky fibers, particularly when observing bright science
targets.

A more serious issue with dedicated sky fibers is their limited
offsets from the center of the science portion of the integral field
unit. The sky fibers for Densepak and Sparsepak are offset 60\arcsec\
and 70\arcsec\ from the center of the science field. For SAURON this
is increased to 154\arcsec\ yet this size of offset can prove
constraining for nearby galaxy work. For a galaxy like M87, whose
half-light radius is $\sim 100$\arcsec\ and possibly larger
\citep{kor09}, the dedicated sky fibers are still collecting a
significant amount of light from the galaxy itself, thus reducing the
final S/N. It is clear that for M87, estimates of the sky at $\sim
150\arcsec$ from the galaxy center will lead to subtraction of some
level of galaxy light. How strong this effect is depends critically on
how far out from the galaxy center the sky estimates come from.

With exposure times of 5 minutes for the sky nods and 20 minutes for
the science frames, weights are given to the sky frames. The weighted
sky frames are then summed to produce an estimate of the night sky. If
the night sky did not evolve over 30 minutes, then the best sky
estimate would come from weighting each neighboring sky nod by 2 and
summing them. However, the sky can evolve on time scales less than 30
minutes. To account for this, 25 different sky estimates are made,
each created by giving different weights to the neighboring sky
nods. These frames are then sent through all of the reductions
independently. Five different weights are used  for each sky nod,
ranging between 1.75 to 2.25 in 0.125 increments. With each sky nod
receiving five different weights, the various combinations of sky nods
lead to the $5^2 = 25$ estimates of the night sky. Once the remaining
data reduction is complete we have 25 versions of each science
frame. This range allows us to analyze, in a very direct way, both the
best sky to subtract from each science frame, and the influence of our
sky subtraction on the final stellar kinematics. We describe here how
the individual sky frames are created from the sky nods and then
discuss how the best sky to subtract from each science frame is
selected from the 25 options.

For each of the 25 scaled sky frames bright continuum objects and
cosmic rays are identified as $3 \sigma$ outliers above the median and
masked. As a $3 \sigma$ cut may not catch low level sources, a 51
fiber-wide boxcar is run over the frame. A boxcar of this size
corresponds to smoothing over a $107\arcsec \times 21\arcsec$ region
of the sky, so even faint, extended sources are removed. From this
frame the sky is modeled by the same method used to model the twilight
sky during the creation of the flat-field. The principle difference is
that rather than modeling the sky to divide out of the frame, the sky
model is what we are after.

The first step in determining the best level of sky to subtract is a
visual inspection of the quality of subtraction of the night sky
lines. However, the determination can not be made based solely on
these lines as they evolve on very short time scales and independently
of the thermal background that most strongly effects our estimates of
the stellar kinematics. The second step is to conduct a preliminary
fit of a single template star (HD20893) to the data. We outline the
steps here, leaving the details of this fitting method to \S
\ref{extract}. First, for each of the 25 data frames, a set of fibers
seeing a moderate level of galaxy light are selected and combined to
form a single spectrum. The exact number of fibers and amount of
galaxy light isn't critical as the final determination comes from a
relative comparison of the results. In fitting the data with a
template star, a convolution occurs between the template star and an
assumed line-of-sight velocity dispersion profile (LOSVD), which
accounts for broadening in the spectra due to the temperature of the
galaxy at that location. Normally, a continuum offset for the template
star is allowed to float when conducting the extraction of
kinematics. However, for this step, this value is fixed to avoid
possible degeneracies between this parameter and the level of sky
subtraction. A comparison of the residuals of the fit between each of
the 25 frames and the single template star is used to determine the
best level of sky subtraction. For nearly all frames ($\sim 90\%$) the
best sky subtraction comes from scaling each sky nod by 2 and summing
them. The exception to this occurs primarily with exposures near dawn
or dusk when twilight begins to affect the weighting.

As the sky evolves on time scales shorter than 30 minutes, the
accuracy of our sky subtraction is not perfect and certain spectral
regions remain problematic, particularly around regions of bright sky
emission lines. In order to get a handle on both the location and
severity of these issues, we conduct a visual inspection of the data
by overplotting the sky-subtracted galaxy spectra for a given spatial
bin with each subtracted sky that goes into a given pointing. We show
an example of the results of this inspection in Figure \ref{sky}. In
the top panel, the 12 different sky spectra subtracted from each 20
minute frame are overplotted on the resulting galaxy spectra. At each
pixel the variance is calculated for the 12 sky spectra and plotted in
the second panel. A region of high variance indicates a region where
the night sky evolves substantially between exposures. Notice that the
high variance tends to be, but is not limited to, spectral regions
with bright sky lines. The lower two panels in Figure \ref{sky} show
the final sky subtracted spectra, along with the best-fit stellar
template, for two spatial bins at R~=~13.2\arcsec\ and
222.0\arcsec. The details of this fitting routine are given in \S
\ref{extract}. The gray areas indicate spectral regions excluded from
the stellar template fitting. These regions are excluded when the
template fit to the data is poor due to either issues with sky
subtraction or template mismatch.

With 25 different estimates of the level of sky subtraction we are
in a good position to explore systematics due to either over or
under subtracting the night sky. To do this we select a range of
spatial bins at various radii and S/N. We then take the reductions up
through extraction of the LOSVD and compare the moments of the LOSVDs
from the 25 frames. Although variation between these frames is seen,
it is both random and within the uncertainties of our analysis. This
is particularly true for central regions of the galaxy where our S/N
is high. At larger radii, where the galaxy light is faint, over or
under subtraction of the night sky tends to wash out the signal
entirely rather than introduce systematics.

There is another component of the reductions that further mitigates
error due to inaccurate sky subtraction. The final spectra is, at
minimum, a combination of three separate frames, and up to 15 for the
case of pointing \#1. As the sky subtraction from each 20 minute
exposure is independent, random poor sky subtraction is mitigated by
having many frames. This mitigating effect gains strength for the more
distant pointings, where the number of exposures increases.

\subsection{Further Reductions}

Once the sky subtraction is complete, cosmic rays are located and
masked. To locate cosmic rays, each pixel value is compared to the
pixel values that fall along either the same row or column of the
extracted frame. Comparison with pixel values along the same row
avoids masking continuum sources while comparison along the same
column avoids masking real galaxy emission features that will appear
in neighboring fibers. A pixel found to be a $7 \sigma$ outlier in
this comparison is masked, as well as all neighboring pixels. Any low
level cosmic rays not masked in this step are rejected when the
spectra from different exposures and fibers is combined.

Next, fibers containing either foreground or background objects are
located and masked. These fibers are identified by taking the median
of the flux in each fiber and plotting these values against the
position on the galaxy. As the median is taken over 5 rows $\times$
2048 pixels = 10,240 values any residual cosmic rays or emission
features will not influence this map. Foreground stars and background
objects are located as outliers from the smooth continuum of the
galaxy and masked. Although these objects will fall onto the same
fibers for the same pointing, each science frame is inspected
individually. We find this frame-by-frame check is necessary as
transient objects, most notably satellites, can swamp an entire row of
fibers.

Low level background sources remain a concern for the most distant
pointings where the surface brightness of M87 may approach the level
of these sources. As the final spectra for the most distant bins comes
from a biweight combination of between 30 to 70 fibers, and even a
large background source will fall into just a few of our large
fibers, any low level source will be rejected by the final biweight
combination.

Each fiber for each night has a unique wavelength solution. Therefore,
a linear interpolation is required before combining spectra from
different fibers. The fiber cross-dispersion profile shape in each
science frame is removed via division by the flat, and so the spectra
from each of the 5 rows of a fiber is used in the biweight combination
independently. In the case of pointing \#3, where we have 3 science
exposures, a minimum of 15 estimates of the spectra go into the
biweight (1 fiber $\times$ 3 exposures $\times$ 5 rows). The biweight
estimator has been shown to be robust for samples smaller than 15 (See
\citet{bee90} for references). For our largest spatial bin, 72 fibers
$\times$ 15 exposures $\times$ 5 rows = 5,400 estimates are sent into
the biweight. Once this step is complete we are left with the 88
VIRUS-P spectra presented in this work.

\clearpage

\begin{figure*}
\centerline{\psfig{file=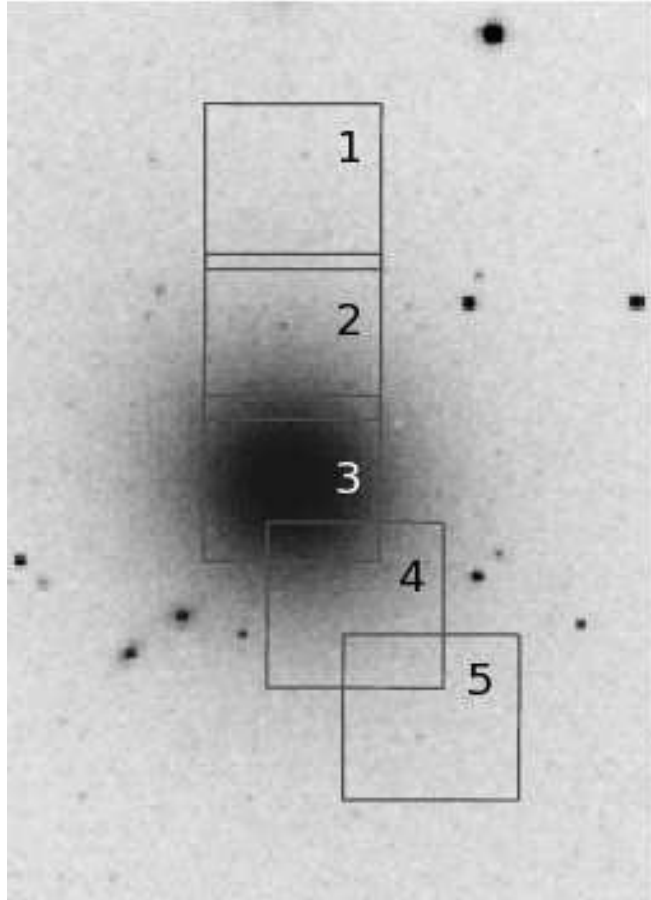,width=8.7cm,height=12.0cm,angle=0}}
\vskip 10pt
\figcaption[M87.ps]{An SDSS image of M87 showing the positions of the
  5 VIRUS-P pointings. Each $107\arcsec \times 107\arcsec$ box consists
  of a hexagonal array of 246 optical fibers (see Figure
  \ref{fiber_pos}). The total exposure time for each pointing is given
  in Table 1. North is up and East is to the right. 
\label{M87pic}}
\end{figure*}

\begin{deluxetable}{cccrr}
\tablewidth{0pt}
\tablecaption{Exposure Times for M87 Pointings
\label{exptime}}
\tablehead{
\colhead{} & \colhead{Exposure} & \colhead{Observation} &
\colhead{R$_{min}$} & \colhead{R$_{max}$}
\\
\colhead{Pointing} & \colhead{Time (min)} & \colhead{Date} & 
\colhead{(\arcsec)} & \colhead{(\arcsec)}
}
\startdata
1  &  180  &  01-08   &  130.0  & 238.0\\
1  &  120  &  02-08   &  130.0  & 238.0\\
2  &  100  &  01-08   &   45.0  & 140.0\\
3  &   60  &  02-09   &    0.0  &  73.0\\
4  &  120  &  02-08   &   43.0  & 136.0\\
5  &  240  &  02-08   &  127.0  & 203.0\\
\tablecomments{The exposure times, date of observation, and radial
  positions of the 5 VIRUS-P pointings on M87. The exposure times
  quoted are the total science exposures included in the final VIRUS-P
  data. Ten of the 51 exposures taken were withheld from the
  reductions based on analysis of the S/N of the resulting
  spectra. Sky nod exposure time is not included in these totals. All
  observing conditions were good to photometric, with typical seeing
  values of 1.5\arcsec. These data were all taken within $\pm3$ days
  of the new moon.}
\end{deluxetable}

\clearpage

\begin{figure*}
\centerline{\psfig{file=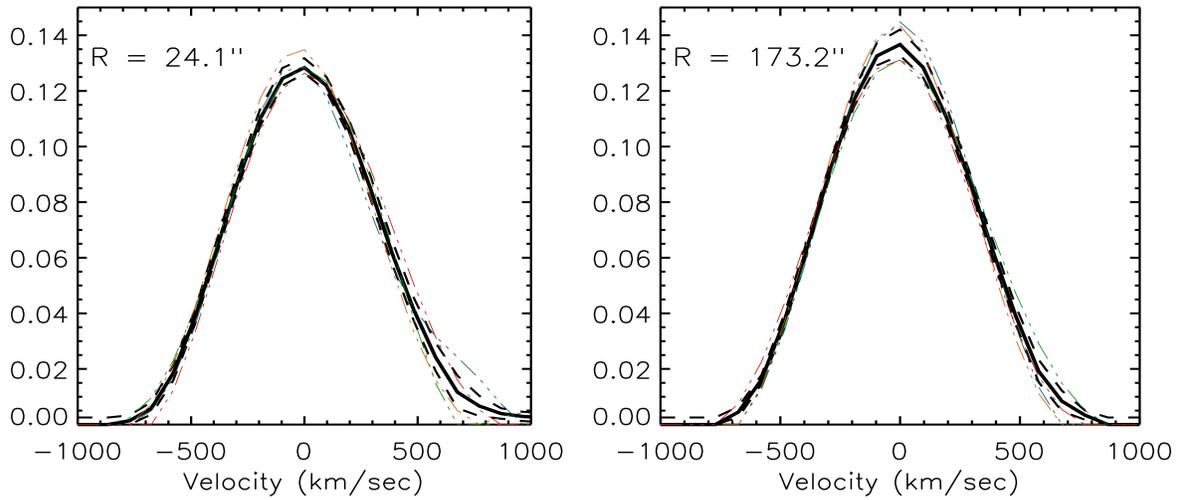,width=16.0cm,height=7.0cm}}
\vskip 10pt

\figcaption[losvd.ps]{Two LOSVDs from spatial bins at R~=~24.1\arcsec\
  (left) and R~=~173.2\arcsec\ (right), plotted with their
  uncertainties. Overplotted with  lighter colored lines are the
  LOSVDs from the four wavelength regions used to determine the final
  LOSVD. Seventy-nine of the 80 final modeling LOSVDs are constructed
  from 4 of the LOSVDs determined from the 4 spectral regions shown in
  Table \ref{regions}. For one spatial bin the iron region
  (5300--5850~\AA) proved a poor fit and was withheld from the final
  LOSVD.
\label{losvdboth}}
\end{figure*}

\begin{deluxetable}{cccrr}
  \tablewidth{0pt}
  \tablecaption{VIRUS-P Spectral Regions
    \label{regions}}
  \tablehead{
    \colhead{Wavelength Range (\AA)} & \colhead{Spectral Features}}
  \startdata
  3650--4050  &  Ca H, Ca K\\
  4195--4585  &  G-band    \\
  4455--4945  &  H-beta    \\
  4930--5545  &  MgI b     \\
  5300--5850  &  Iron      \\
  \tablecomments{The 5 spectral regions chosen for extraction of 
    the best-fit LOSVD for each spatial bin. The calcium H \&\ K
    region (3650--4050~\AA) is not used in the determination of the
    final LOSVD due to a systematic offset in the measured velocity
    dispersion as compared to the other spectral regions. This
    systematic is likely due to issues with the continuum normalization
    over the blue region of the spectra (see Footnote \ref{ftnt1} and
    \S \ref{alpha} for further discussion).}
\end{deluxetable}

\begin{deluxetable}{lllr}
  \tablewidth{0pt}
  \tablecaption{Indo-US Template Stars
    \label{tlist}}
  \tablehead{
    \colhead{ID} & \colhead{Type} & \colhead{V} & \colhead{[Fe/H]}}
  \startdata
  HD 50420  &  A7III &  6.16  &  0.30 \\
  HD 78362  &  F5III &  4.65  &  0.52 \\
  HD 5015   &  F8V   &  4.82  &  0.00 \\
  HD 693    &  F5V   &  4.89  & -0.38 \\
  HD 39833  &  G0III &  7.66  &  0.04 \\
  HD 161797 &  G5IV  &  3.41  &  0.16 \\
  HD 199960 &  G1V   &  6.21  &  0.11 \\
  HD 17820  &  G5V   &  8.38  & -0.69 \\
  HD 20893  &  K3III &  5.09  & -0.13 \\
  HD 6734   &  K0IV  &  6.46  & -0.25 \\
  HD 92588  &  K1IV  &  6.26  & -0.10 \\
  HD 130025 &  K0V   &  6.16  & -0.19 \\
  \tablecomments{The template stars used in the determination of the
    best-fit LOSVD. These 12 stars were selected from an initial list
    of 40 stars based on a minimization of the fitting residuals
    during the kinematic extraction.}
\end{deluxetable}

\begin{figure*}
\centerline{\psfig{file=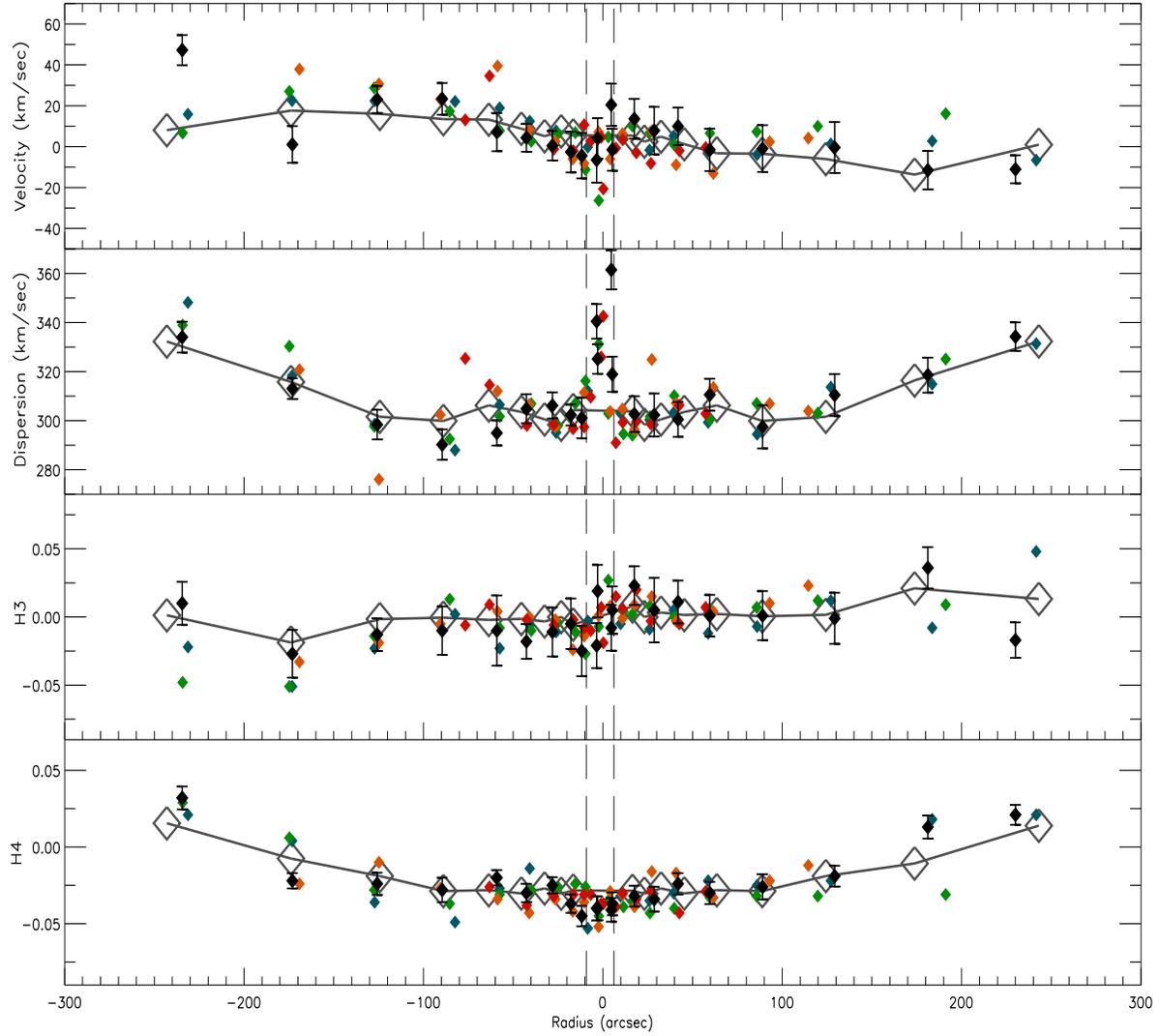,height=15.0cm,width=17.0cm}}
\vskip 10pt
\figcaption[M87_moments.ps]{The first 4 moments of the Gauss-Hermite
  expansion of the 88 VIRUS-P LOSVDs. The filled diamonds show the
  data at different angular bins. The black diamonds are for the major
  axis, followed by blue, green, orange, and with red along the minor
  axis. The dashed vertical lines near the center indicate the region
  where VIRUS-P data is not used and SAURON kinematics are employed in
  the modeling. The open diamonds, connected by a line, plot the
  moments, averaged over the angular bins, from the best-fit
  logarithmic dark halo model.
\label{moments}}
\end{figure*}

\begin{figure*}
  \centerline{\psfig{file=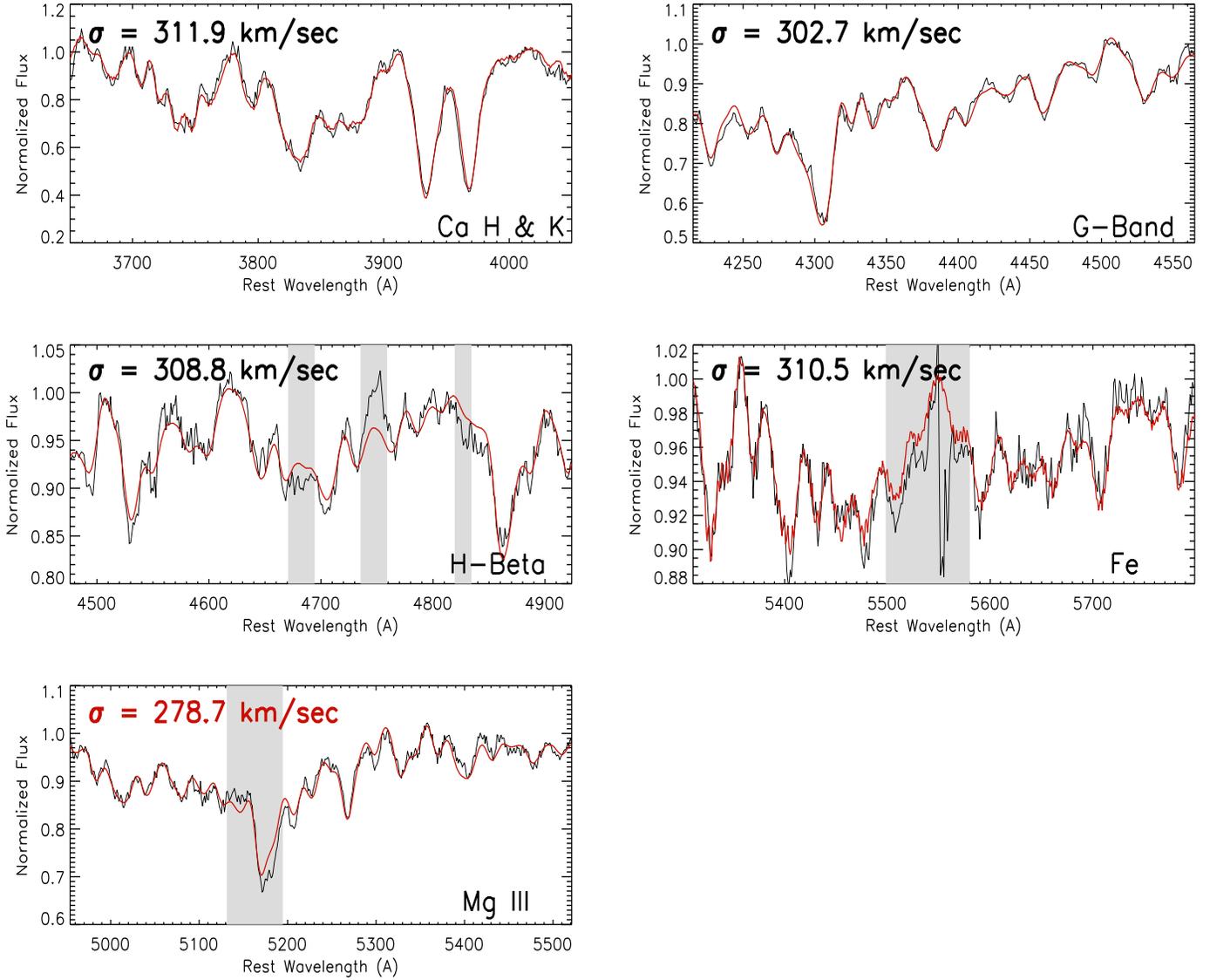,width=19cm,height=15.6cm}}
  \vskip 10pt
  \figcaption[compare.ps]{VIRUS-P data for 5 spectral regions from the
  spatial bin at R~=~24.1\arcsec. The black line plots the data and the
  red line plots the best-fit stellar template. The shaded gray
  regions are withheld from the kinematic extraction as discussed in \S
  \ref{extract}. The velocity dispersions measured for each of the 5
  spectral regions are shown in the upper-left of each panel. The
  systematic offset between the Mg~$b$ region ($\sigma = 278.7$~\kms)
  and the other 4 spectral regions (with a mean of $\sigma =
  308.5$~\kms) is clear. The Ca H~+~K region, while initially withheld
  from the dynamical modeling due to a large systematic offset seen in
  its calculated Gauss-Hermite moments, is included here. The offset
  seen in Ca H~+~K was due to issues of continuum normalization. Since
  completion of the dynamical modeling this issue was resolved and can
  now be included in this comparison.
  \label{regcomp}}
\end{figure*}

\begin{figure*}
\centerline{\psfig{file=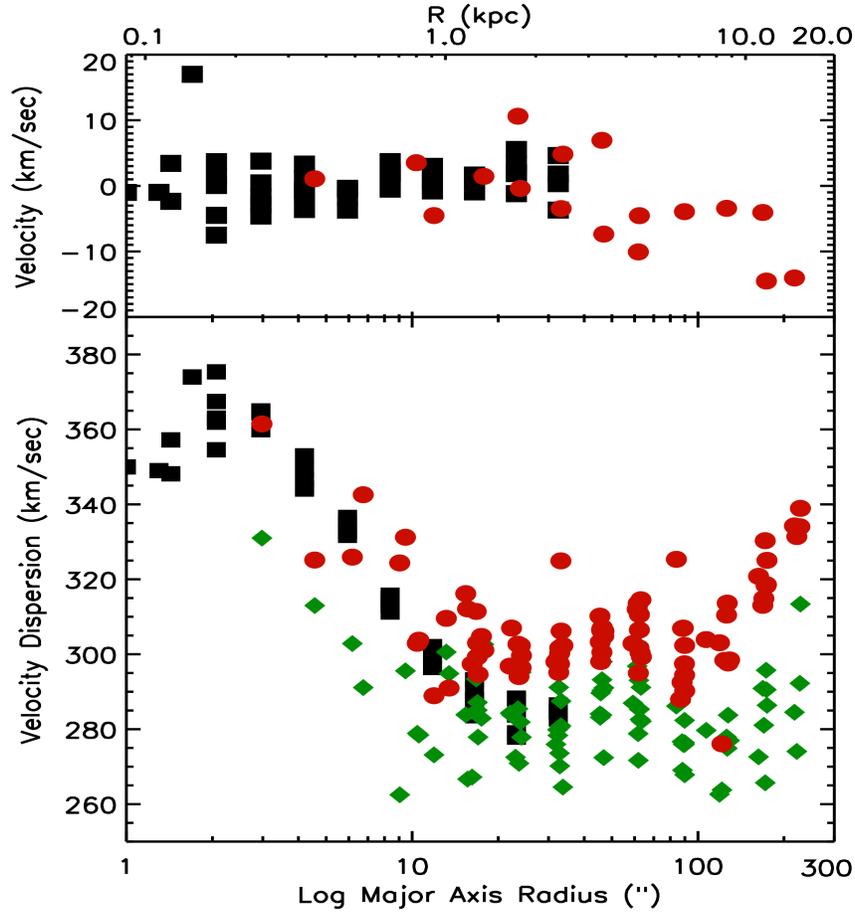,width=11.8cm,height=12.6cm}}
\vskip 10pt
\figcaption[SAURONvVP_log.ps]{Velocity (top) and velocity dispersion
  (bottom) measurements from SAURON and VIRUS-P. Black squares plot
  the SAURON data and red circles the VIRUS-P data. The green diamonds
  show the velocity dispersion measured with VIRUS-P over the Mg~$b$
  region. These velocity dispersion values are offset by  $\sim
  20$~\kms\ from the velocity dispersion values calculated when
  combining spectral regions. With a wavelength range covering the
  H-beta and Mg~$b$ spectral regions, the SAURON spectral range is
  similar to the VIRUS-P Mg~$b$ region. The agreement between the
  SAURON and VIRUS-P velocity dispersion values over this region
  (277.0~\kms\ and 281.8~\kms\ respectively) is within our
  uncertainties.
\label{offset}}
\end{figure*}

\begin{figure*}
  \centerline{\psfig{file=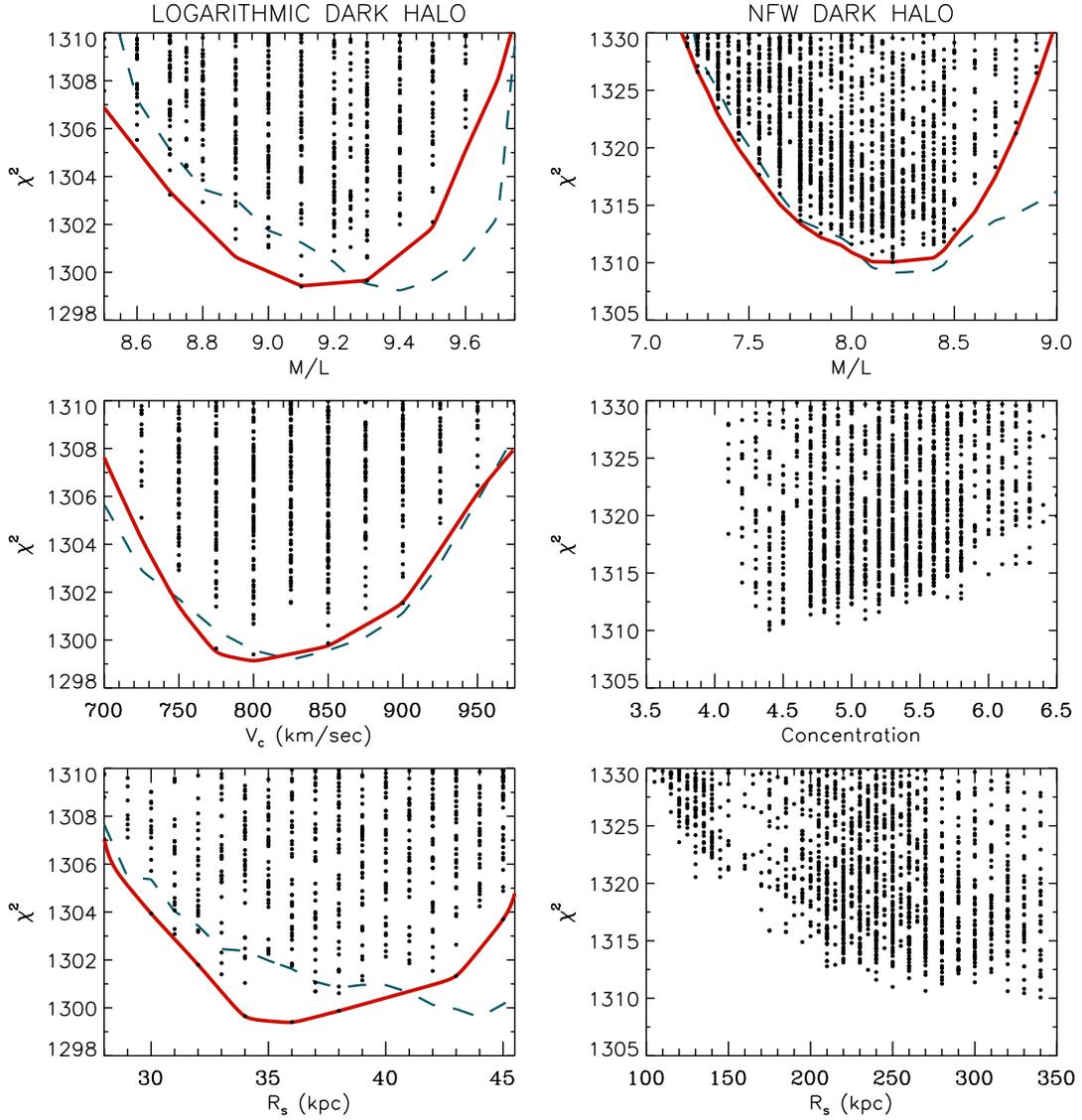,width=22.0cm,height=15.4cm}}
  \vskip 20pt
  \figcaption[chi1.ps]{The $\chi^2$ values (stars + GCs) vs. the three
  modeling parameters for a logarithmic dark matter halo (left) and
  NFW halo (right). Each black dot is the $\chi^2$ value from a
  single model. To highlight the variation at the $\chi^2$ minimum,
  only a few hundred of the 1000s of models run are shown. A smoothed
  spline fit to the minimum $\chi^2$ values (plotted in red) gives
  us our 68\%\ ($\Delta \chi^2 \le 1$) and 95\%\ ($\Delta \chi^2 \le
  2$) confidence bands. The dashed blue line plots the $\chi^2$
  minimum values for the stars. An additive shift of 41.5 has been
  made to the stellar values. As the shift is additive, the relative
  $\chi^2$ values are preserved. The NFW halo results show the lack
  of constraint on the NFW scale radius parameter (lower right
  panel). We do not show similar spline fits to the NFW halo due to
  the unconstrained nature of the model. We note that while we do not
  constrain the NFW dark halo parameters, the constraint on the
  stellar M/L is very robust. The $\Delta$M/L of 1.1 between the
  logarithmic and NFW models is due in large part to the difference in
  inner slope of the assumed dark halo parameters. 
\label{chi1}}
\end{figure*}

\begin{figure*}
  \centerline{\psfig{file=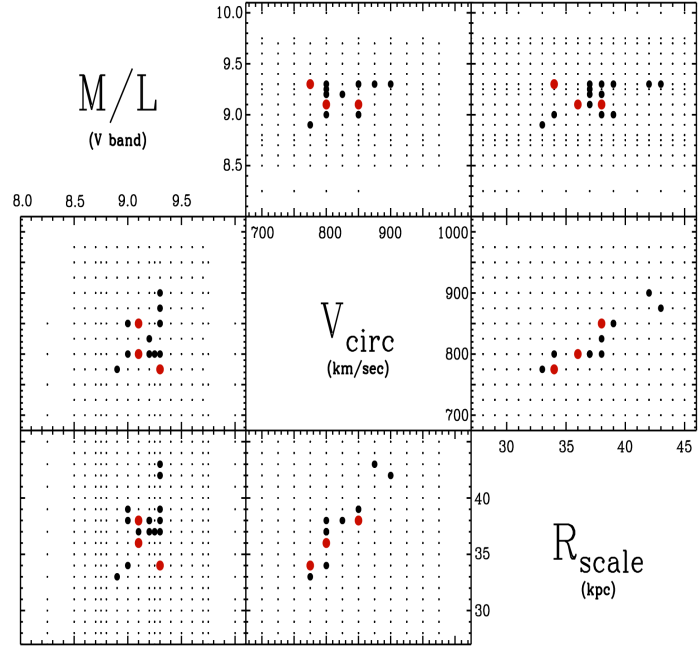,width=10.0cm,height=9.5cm}}
  \vskip 10pt
\figcaption[chi2.ps]{Plots of the $\chi^2$ minimums of the 3
  parameters plotted against one another for the logarithmic dark matter
  halo. The $\chi^2$ range shown is the same as in Figure \ref{chi1}
  (left half). The small black dots show individual models that lie
  near the $\chi^2$ minimum. The larger black dots show models that
  fall within the 95\%\ confidence band ($\Delta \chi^2 \le 2$) while
  the larger red dots show models within the 68\%\ confidence band
  ($\Delta \chi^2 \le 1$). The modeling degeneracy between the dark
  and luminous matter, as discussed in GT09, is clearly seen in the
  correlation between the stellar M/L and the two dark matter halo
  parameters.
    \label{chi2}}
\end{figure*}

\begin{figure*}
  \centerline{\psfig{file=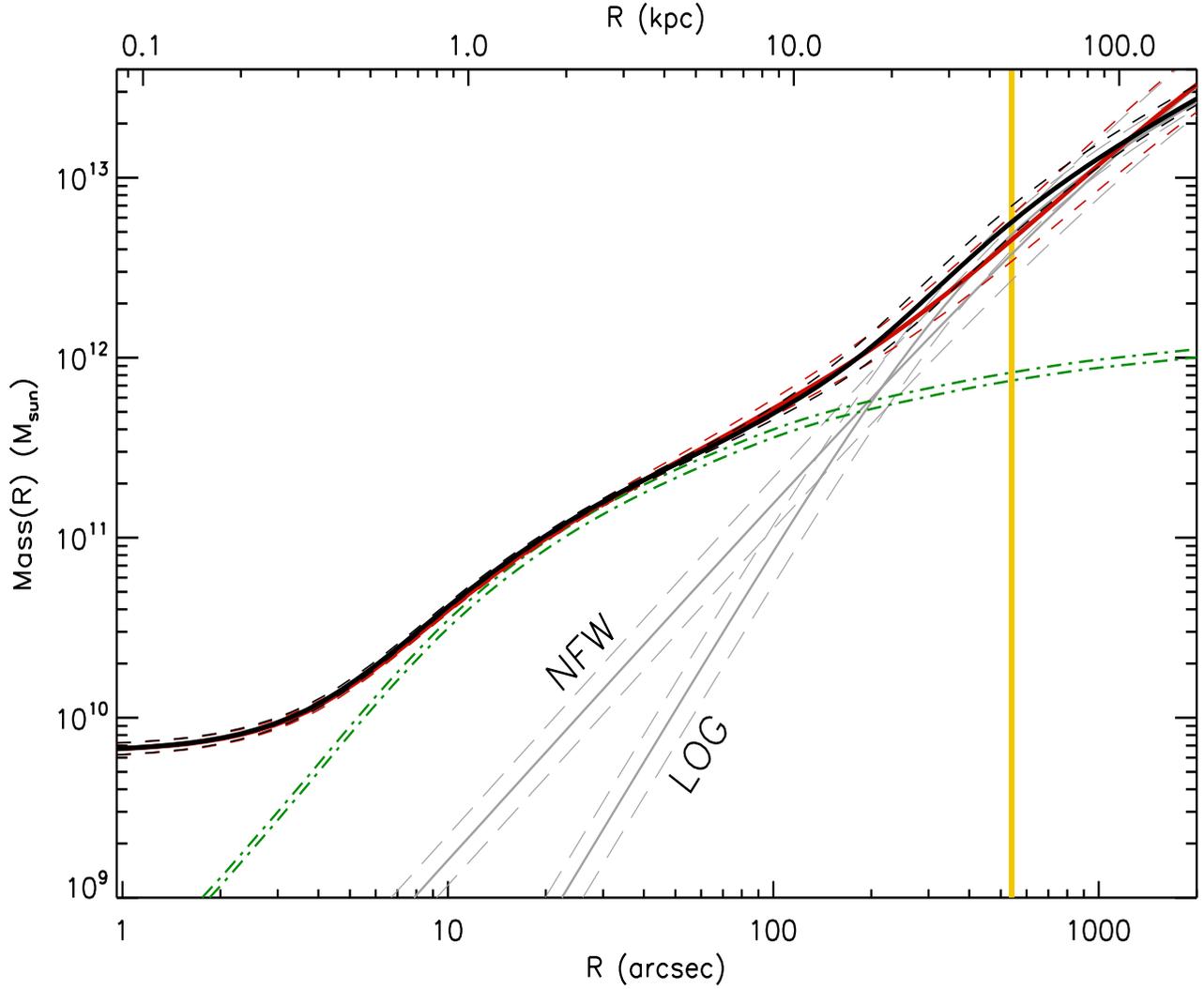,width=18cm,height=14.5cm}}
  \vskip 10pt
\figcaption[total_mass.ps]{Total enclosed mass as a function of
    radius. The solid, black line indicates total enclosed mass for
  our best-fit logarithmic model. The NFW enclosed mass profile is
  plotted in red. Both of these enclosed mass models are plotted with
  uncertainties, which are the min/max values for the $4^2 = 16$
  dynamical models that explore the parameter limits of our 68\%\
  confidence bands. The green lines plot stellar mass for both
  models (with uncertainties less than the thickness of the line) and
  the light gray lines, with uncertainties, indicate the two assumed
  dark matter distributions. Note the total enclosed mass does not go
  to zero with radius due to inclusion of a $6.4 \times 10^9~M_\odot$
  mass black hole. Modeling results beyond our last data point,
  indicated by the vertical yellow line, are not constrained by the
  data, and are therefore suspect.
    \label{mass}}
\end{figure*}

\begin{figure*}
  \centerline{\psfig{file=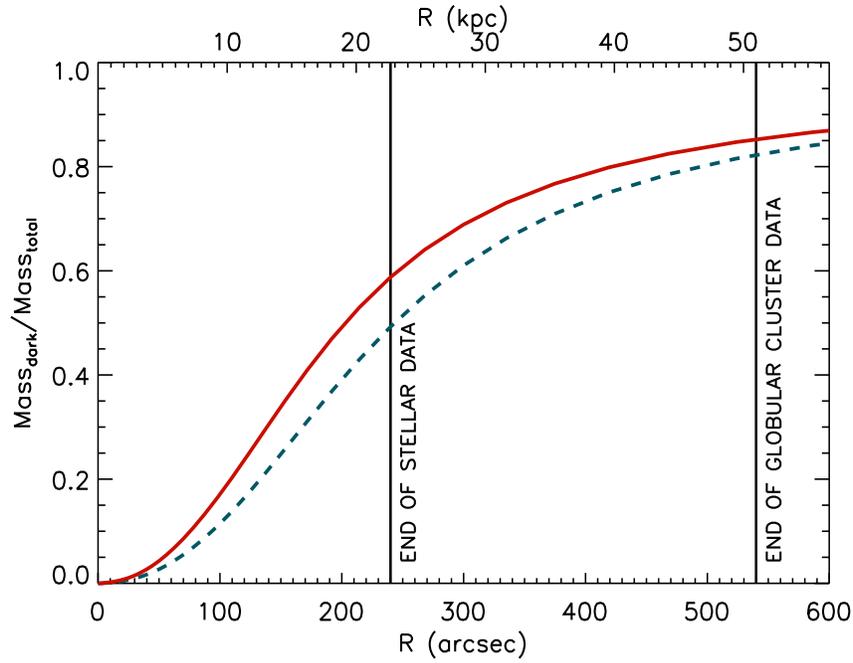,width=12.0cm,height=9.0cm}}
\vskip 10pt
\figcaption[DM_total_ratio.ps]{The enclosed dark matter fraction as a
  function of radius for a logarithmic halo. The red (solid) line
  shows the best-fit $\chi^2$ model for both stars and globular
  clusters (i.e. $\chi^2 = \chi^2_{\text{stars}} +
  \chi^2_{\text{GC}}$). The blue (dashed) line shows the best-fit
  dynamical model based on the $\chi^2$ value for stars only
  (i.e. $\chi^2 = \chi^2_{\text{stars}}$). This figure indicates the
  degree to which the large radii GCs influence the dark matter
  fraction at all radii.
\label{mass2}}
\end{figure*}

\begin{deluxetable*}{ccccccc}[t]
  \tablewidth{0pt}
  \tablecaption{Literature Mass Comparison
    \label{masscompare}}
  \tablehead{
    \colhead{Reference} & \colhead{Method} & \colhead{Symbol} & \colhead{Radius} & \colhead{Literature} & \colhead{Logarithmic} & \colhead{NFW}
    \\
    \colhead{(year)} & \colhead{} & \colhead{} & \colhead{(arcsec)} & \colhead{$10^{12}M_\odot$} & \colhead{$10^{12}M_\odot$} & \colhead{$10^{12}M_\odot$}}
  \startdata
                             &                    &         &       &                         &                        &                        \\
  Neito \& Monnet (1984)     & Empirical          & circle  &  490  &  $5.20                $ & $4.90^{+1.14}_{-0.82}$ & $3.90^{+1.39}_{-0.91}$ \\
  Brandt \& Roosen (1969)    & Empirical          & diamond &  84   &  $2.7^{+1.4}_{-1.4}$    & $4.07^{+0.28}_{-0.25}$ & $4.30^{+0.49}_{-0.34}$ \\
  Poveda (1961)              & Empirical          & square  &  84   &  $1.4^{+3.4}_{-0.4}$    & $0.42^{+0.03}_{-0.03}$ & $0.44^{+0.05}_{-0.03}$ \\
  Fabricant et al. (1983)    & X-rays             & circle  &  1336 &  $15.5^{+3.5}_{-3.5}$   & $17.9^{+3.3}_{-1.6}$   & $18.3^{+8.9}_{-5.2}$   \\
  Huchra \& Brodie (1987)    & GC kinematics      & circle  &  248  &  $6.1^{+2.2}_{-2.2}$    & $1.61^{+0.35}_{-0.29}$ & $1.47^{+0.37}_{-0.26}$ \\
  Mould et al. (1987)        & GC kinematics      & diamond &  200  &  $0.90^{+0.15}_{-0.15}$ & $1.15^{+0.22}_{-0.18}$ & $1.12^{+0.25}_{-0.17}$ \\
  Sargent et al. (1978)      & Stellar kinematics & diamond &  80   &  $0.19^{+0.10}_{-0.20}$ & $0.39^{+0.03}_{-0.02}$ & $0.41^{+0.05}_{-0.03}$ \\
  Sargent et al. (1978)      & Stellar kinematics & diamond &  47   &  $0.14^{+0.05}_{-0.05}$ & $0.24^{+0.01}_{-0.01}$ & $0.24^{+0.02}_{-0.01}$ \\
  Tsai (1996)                & X-rays             & diamond &  266  &  $2.20                $ & $1.80^{+0.41}_{-0.33}$ & $1.61^{+0.42}_{-0.29}$ \\
  Merritt \& Tremblay (1993) & GC kinematics      & square  &  603  &  $6.0^{+4.0}_{-1.0}$    & $6.66^{+1.50}_{-0.99}$ & $5.36^{+2.07}_{-1.32}$ \\
  Matsushita et al. (2002)   & X-rays             & square  &  113  &  $0.43^{+1.0}_{-1.0}$   & $5.58^{+0.58}_{-0.50}$ & $5.92^{+0.85}_{-0.60}$ \\
  Matsushita et al. (2002)   & X-rays             & square  &  226  &  $1.0^{+1.0}_{-1.0}$    & $1.39^{+0.29}_{-0.24}$ & $1.30^{+0.32}_{-0.21}$ \\
  Matsushita et al. (2002)   & X-rays             & square  &  340  &  $2.0^{+1.0}_{-1.9}$    & $2.71^{+0.65}_{-0.50}$ & $2.26^{+0.69}_{-0.46}$ \\
  Wu \& Tremaine (2006)      & GC  kinematics     & circle  &  406  &  $2.4^{+0.6}_{-0.6}$    & $3.64^{+0.87}_{-0.65}$ & $2.93^{+0.97}_{-0.65}$ \\
  \tablecomments{Enclosed mass values from the literature. C1)
    Reference. C2) Method employed to determine enclosed mass. C3)
    Symbol used to plot the data in Figure \ref{others}. C4) Radial
    distance from the center of the galaxy, scaled to the distance
    assumed in this work (R~=~17.9~Mpc). C5) Literature value of
    enclosed mass at the radial position in C4. The uncertainty is
    quoted, where available. C6) Enclosed mass from the best-fit
    logarithmic halo model from this work. C7) Enclosed mass from the
    best-fit NFW halo model from this work.}
\end{deluxetable*}

\begin{figure*}
  \centerline{\psfig{file=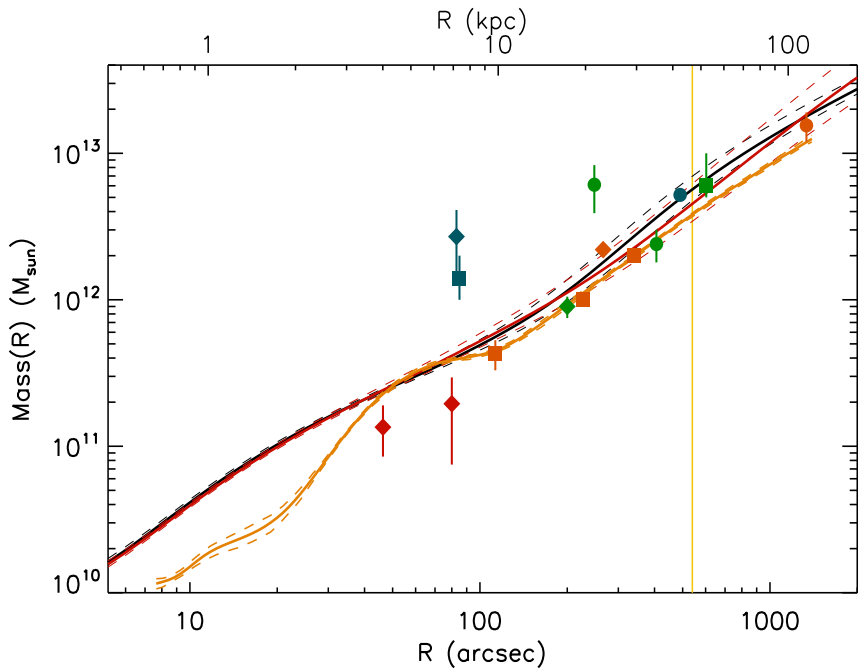,width=12.0cm,height=9.0cm}}
\vskip 10pt
\figcaption[total_mass2.ps]{A comparison of total enclosed mass from
  the literature. The symbols are explained in Table \ref{masscompare}.
  The color of the symbols indicates the method employed to make the
  mass determination. Blue: empirical, Green: GC kinematics, Red:
  Stellar kinematics, Orange: X-rays. The red, black and yellow lines
  are described in Figure \ref{mass}.
\label{others}}
\end{figure*}

\begin{figure*}
  \centerline{\psfig{file=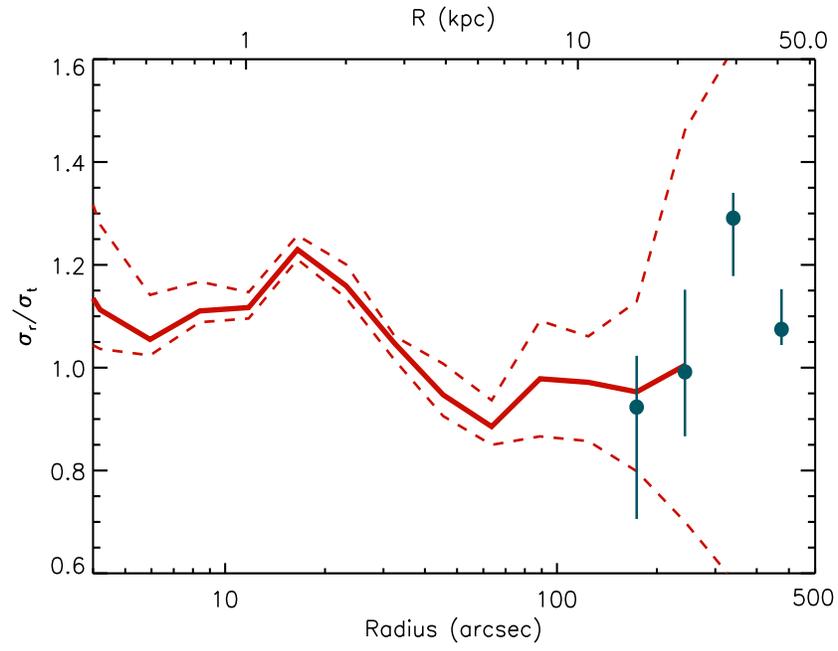,width=12.0cm,height=9.0cm}}
\vskip 10pt
\figcaption[intmom.ps]{The ratio of the radial velocity anisotropy to
  the tangential anisotropy for both the stars (red lines) and GC
  (blue dots).
\label{intmom}}
\end{figure*}

\end{document}